# Studies On Falling Ball Viscometry

Amit Vikram Singh[1], Lavanjay Sharma[2], and Pinaki Gupta-Bhaya[1]

Departments of Chemistry[1], Material Science Programme[2], Indian Institute of Technology Kanpur, Kanpur 208016

India

**Abstract:** A new method of accurate calculation of the coefficient of viscosity of a test liquid from experimentally measured terminal velocity of a ball falling in the test liquid contained in a narrow tube is described. The calculation requires the value of a multiplicative correction factor to the apparent coefficient of viscosity calculated by substitution of terminal velocity of the falling ball in Stokes formula. This correction factor, the so-called viscosity ratio, a measure of deviation from Stokes limit, arises from non-vanishing values of the Reynolds number and the ball/tube radius ratio. The method, valid over a very wide range of Reynolds number, is based on the recognition of a relationship between two measures of wall effect, the more widely investigated velocity ratio, defined as the ratio of terminal velocity in a confined medium to that in a boundless medium and viscosity ratio. The calculation uses two recently published correlation formulae based on extensive experimental results on terminal velocity of a falling ball. The first formula relates velocity ratio to Reynolds number and ball-tube radius ratio. The second formula gives an expression of the ratio of the drag force actually sensed by the ball falling in an infinite medium to that in the Stokes limit as a function of Reynolds number alone. It is shown that appropriate use of this correction factor extends the utility of the technique of falling ball viscometry beyond the very low Reynolds number 'creepy flow' regime, to which its application is presently restricted. Issues related to accuracy are examined by use of our own measurements of the terminal velocity of a falling ball in a narrow tube and that of published literature reports, on liquids of known viscosity coefficient.

**1. Introduction**

**1.1** Evaluation of hydrodynamic forces on a rigid body in relative motion in a fluid has been of interest for a very long time (Clift et. al [1], Happel and Brenner [2], Kim and Karrila 2005 [3]) and also recently (Leach 2009 [4]). A falling spherical ball, which senses this force, has been used as a probe to study fluid properties. Measurement of terminal velocity ($V_t$) of a ball falling in a viscous fluid enclosed in a narrow tube provides a method for determination of the coefficient of viscosity ($\mu$) of the test liquid. This simple, yet accurate technique, in use for a long time, is of considerable recent interest (Kahle et. al. [5], Kaiser et. al. [6], Brizard et. al. [7], Feng et. al. [8], Ma et. al. [9]). A falling ball viscometer is commercially available and has been used for testing petroleum products, pharmaceutical beverages, silicate glass and food products.

In addition to viscometry, study of a falling ball is important in several engineering domains which involve multiphase flows e.g., sedimentation, improvement of combustion, minimization of erosion by droplets in large turbines, hydrodynamic chromatography, membrane transport, hydraulic and pneumatic transport of coarse particles in pipes, effects that utilize electric fields to enhance transport phenomena and separations in multiphase systems (Kaji et. al. [10], Scott and Wham [11], Ptasinski and Kerkhof [12]).

Motion of a falling ball in a liquid contained in a narrow tube, apart from viscometry and other practical applications, is interesting in its own right. Eccentric fall, horizontal wall forces, and accelerated pre-steady state fall are a few examples of many interesting aspects of the physics of a falling ball (Happel and Brenner [2], Mordant and Pinton [13], Tozeren [14], Rubinow and Keller [15], Shinohara and Hashimoto [16], Ambari et. al. [17], Humphrey and Murata [18], Becker and Mc Kinley [19], Bougas and Stamatoudis [20], Feng et. al. [21], Changfu et. al. [22]).

Falling ball viscometry assumes importance in the study of non-Newtonian fluids, an area of considerable recent interest. The use of more conventional viscometers, e.g., capillary or rotary, for zero-shear rate viscosity measurement of non-Newtonian fluids is error-prone. Measurements at low shear rate are problematic in these instruments and extrapolation to zero shear rate is ambiguous. Falling ball viscometer is superior in this regard and considerable amount of work has been done in application of falling ball viscometry to measurement of viscosity of non-Newtonian fluids, both experimental and in respect of techniques of extrapolation to zero shear stress (Kaiser et. al. [6], Williams [23], Sutterby [24], Turian [25], Caswell [26], Cygan and Caswell [27], Subbaraman et. al. [28], Chhabra and Uhlherr [29], Barnes [30], [31]).

**1.2 Velocity ratio:** A ball falling under the force of gravity in a fluid attains a terminal steady velocity when the frictional (drag) force ($F_D$) exactly balances the sum of oppositely directed force of gravity and force due to buoyancy. This sum and therefore, the drag force on the ball in its state of steady fall, are determined by the ball-fluid combination alone. Its magnitude is entirely independent of the presence or absence of the walls of the tube in the vicinity of the falling ball. Drag force is a function of ball velocity. Although the numerical value of $F_D$ is independent of the proximity of the ball with respect to the wall, its functional relation with $V_t$ is not. A ball liquid combination with a given $F_D$ will show a tube diameter dependent $V_t$. The magnitude of $V_t$ in an unbounded medium is determined entirely by the ball fluid combination and that in a confined medium is determined, in addition, by the location of the falling ball with respect to the wall. The value of $V_t$ in an infinite medium, designated $V_\infty$ and the dimensionless Reynolds number $Re_\infty$ (Eq. 2) that quantifies inertial effect and is proportional to $V_\infty$ are characteristic parameters of the ball-liquid combination under test. The parameter that quantifies proximity of the ball to the wall for centerline fall is the ratio of ball diameter to tube diameter, designated $\lambda$. The modification of terminal velocity on confinement quantified by the ratio $V_t/V_\infty$ is a function of $\lambda$, the functional dependence being parametrically dependent on the ball-liquid specific parameter $Re_\infty$. There exists extensive literature on measurements and parameterization of $V_t/V_\infty$ as a function of $Re_\infty$ and $\lambda$ (Chhabra [32], DiFelice [33], Kehlenbeck and DiFelice [34], Francis [35], Fidleris and Whitmore [36]; henceforth referred to as F&W), McNown [37]).

**1.3 Viscosity ratio: Falling ball viscometry:** In the limiting case of a boundless fluid medium ($\lambda \to 0$) and a negligible inertial effect ($Re_\infty \ll 1$) Stokes equation specifies the dependence of $F_D$ on $V_t$, a linear function

$$F_D = 6\pi r \mu_\infty V_t \tag{1}$$

$\mu_\infty$ denotes coefficient of dynamic viscosity of the fluid, r is ball radius and $Re_\infty$ is defined by

$$Re_\infty = \frac{\rho_l V_\infty d}{\mu_\infty} \tag{2}$$

where $\rho_l$ is liquid density and d is ball diameter

If one makes the unjustified assumption that Stokes equation holds in a narrow tube, one can calculate $\mu_\infty$ from experimentally determined $V_t$ with use of the expression of $F_D$ in terms of physical parameters of the ball and the liquid. Deviation from Stokes equation is accommodated by defining an apparent so-called Stokes coefficient of viscosity $\mu_S$ in the structure of Stokes equation. The ratio $\mu_S/\mu_\infty$ ($>1$) designated viscosity ratio, is a measure of deviation from Stokes equation. The expression of $\mu_S$ in terms of $V_t$ and physical parameters of the ball and the liquid is given by

$$\mu_S = \frac{2(\rho_b - \rho_l)r^2 g}{9 V_t} \tag{3}$$

where $\rho_b$ is ball density, $\rho_l$ is fluid density, $g$ is acceleration due to gravity, and $r$ is ball radius. One can determine $\mu_S$ for a test liquid from measured $V_t$ and then $\mu_\infty$ from a knowledge of $\mu_S/\mu_\infty$.

Sutterby [38] reports experimental values of $\mu_S/\mu_\infty$ for equally spaced values of $\lambda$ ($0 < \lambda < 0.13$), and closely spaced small values of Reynolds number $Re$ ($0 < Re < 4$) and $Re_S$ ($0 < Re_s < 3$), measures of inertial effects defined as follows

$$Re = \frac{\rho_l V_t d}{\mu_\infty} \tag{4}$$

$$Re_s = \frac{\rho_l V_t d}{\mu_S} \tag{5}$$

Whereas $Re_\infty$ is characteristic of ball-liquid combination alone, $Re$ and $Re_s$ are also determined by $\lambda$ through its effect on $V_t$. The three Reynolds numbers are equal in a boundless medium; $Re$ and $Re_s$ in a narrow tube are less than $Re_\infty$ for the same ball liquid pair. The ratio $\mu_S/\mu_\infty$, shown to be a function of $Re$ and $\lambda$ alone, are calculated at equally spaced values of $Re$ and $\lambda$ by interpolation of extensive experimental data for a wide variety of ball-liquid-tube combination. $\mu_S$ is calculated from experimental values of $V_t$ in a falling ball viscometer and $\mu_\infty$ is measured in an independent viscometer. The lack of information on $\mu_S/\mu_\infty$ as a function of Reynolds number over the whole range is the only limiting factor in the use of falling ball viscometry to determination of absolute value of $\mu_\infty$.

If measurements on $V_t$ for a given test liquid could be restricted to low Reynolds number regime covered by Sutterby's work, falling ball viscometry would be applicable to any liquid. It may however, not be possible to abide by this restriction in all situations. In falling ball experiments one has restriction on ball radius imposed by experimental convenience of position detection and restriction on ball density by the requirement that it must exceed the density of the test liquid as well as commercial availability. With these restrictions $F_D$ varies within a range. The consequence of this restricted range is that low Reynolds number "creepy flow" regime may not always be easy to achieve, particularly for liquids of low viscosity. We consider this issue in sec. 4.1.1.

One way to extend the applicability of falling ball viscometry beyond the range covered by Sutterby [38] is to extend measurements of $\mu_S/\mu_\infty$ that he reports to a wider range of $Re$. This is a tall order. However, extensive data on $V_t/V_\infty$, already exists over the whole range of $\lambda$ and a very wide range of $Re_\infty$. Both of these ratios are measures of deviation from Stokes equation. They are equal only in the limit $Re_\infty \to 0$. The relation between the two ratios can be used for using velocity ratio data for calculation of viscosity ratio, well beyond the 'creepy flow' regime.

The unavailability of viscosity ratio data at large $Re_\infty$ has restricted the use of falling ball viscometry to the 'creepy flow' regime. Several very recent studies on falling ball viscometry (Kahle et. al. [5], Brizard et. al. [7] and Ma et. al. [9]) are also restricted to low $Re_\infty$ regime. We are not aware of its application to larger $Re_\infty$ regime, where $\mu_S/\mu_\infty$ does not equal $V_t/V_\infty$.

**1.4 Force ratio:** We define a third measure of deviation from Stokes limit, the so called force ratio, and discuss its relation to the two measures already defined.

If the conditions for validity of Eq. 1 do not hold, then the expression on R.H.S. of Eq. 1 no longer gives the drag force and functional dependence of $F_D$ on $V_t$ is no longer linear. This expression, as a limiting force still remains a part of description of the phenomena we study. A symbol $F_S$ is assigned to it. It represents Stokes limit for a given terminal velocity. We define

$$F_S = 6\pi r \mu_\infty V_t \qquad (6)$$

If conditions of validity of Stokes equation hold $(F_D/F_S) = 1$.

The deviation of this ratio from unity in an infinite medium arises from a non negligible $Re_\infty$. In a confined medium $\lambda \neq 0$ is an additional source of deviation. Deviation from Stokes limit implies nonlinear functional dependence of $F_D$ on $V_t$ or $V_\infty$ as the case may be. $F_S$ being a linear function, $F_D/F_S$ ( $> 1$) is, in unbounded and in confined medium, a non linear function of $V_t$. The functional dependence of $F_D/F_S$ in an unbounded medium $((F_D/F_S)_{\lambda \to 0})$ on $V_\infty$ has been specified in terms of the dimensionless $Re_\infty$, in the form of a power series using perturbation theory (Proudman and Pearson [39]) and also by parameterization of extensive experimental results on $V_\infty$ (Clift et. al. [1] and Cheng et. al. [40]). In a confined medium $F_D$ is $\lambda$ independent, but $F_S$ being proportional to $V_t$ is not; as a result $(F_D/F_S)_{\lambda \neq 0}$ is a function of $\lambda$ as well as $Re$. This functional dependence has been studied using computer simulation results (Wham et. al. [41]). Purely theoretical evaluation of a limiting force ratio $(F_D/F_S)_{Re \to 0}$ has received attention for more than a century (Happel and Brenner [2]$^a$). The theories have been classified as 'exact' and 'approximate'. Results of the former category are given in the form of tables (Tozeren [14], Haberman [42], Payne and Scherr [43], Coutanceau [44], Bohlin [45]) and those of the latter are given in the form of a compact function of $\lambda$ (Haberman and Sayre [46]).

Number of reports on direct force measurements on suspended spheres to determine $F_D/F_S$ is not large. Ambari et. al. [17], [47] use a magnetic rheometer to accurately measure force ratio at very low Reynolds number ($Re = 10^{-3}$). The results of 'exact' theory are found to be in complete agreement with the experimental results of Ambari et. al. [17]. The agreement between a careful experiment and a rigorous theory verifies both in one shot. Force ratio in the limit $Re \to 0$ equals velocity ratio. The determination of velocity ratio does not require the value of $V_t$. The magnetic ball is held stationary by application of external magnetic force and a fluid-filled tube is moved at a fixed velocity past the stationary ball. The frictional force $F_S$ so generated is measured by measurement of additional external magnetic force required to hold the ball stationary.

A recent work on force ratio also restricted to the creepy flow regime is that of Leach et. al. [4]. They use Faxen formula at low $Re_\infty$ (Happel and Brenner [2]$^b$) to interpret experimental results on translational and rotational drag on optically trapped spherical particles near a wall measured using optical tweezers, remaining within the creepy flow regime. The translational drag to particle movement parallel to a wall, at a location very close to the wall and at substantial distances from the wall (but not at the center of the tube) show impressive agreement with those calculated from Faxen equation. Of the three ratios, only the force ratio does not require a measurement in an infinite medium.

**1.5 Relation between ratios:** In section 3.1 we give two relations, the equality of $(V_t/V_\infty)^{-1}$ and the ratio of $(F_D/F_S)_{\lambda \neq 0}$ to $(F_D/F_S)_{\lambda \to 0}$ and that of $F_D/F_S$ to $\mu_S/\mu_\infty$ and use them to extend falling ball viscometry well beyond the 'creepy flow' regime.

**1.6 Limiting values of ratios:** In the limit $Re_\infty \to 0$, $V_t$ dependence of drag force is linear; all three ratios are equal and are functions of $\lambda$ alone. Highly accurate estimates, theoretical and experimental, are available in this limit (Ambari et. al. [17], Happel and Brenner [2$^a$]). Since $\lambda$ is accurately determined, the correction factors to Stokes equation are easily and accurately determined. In this limit all ball liquid pairs show identical deviation from Stokes equation, their specific physical signatures being absent.

Stokes limit $(F_D/F_S) \to 1$ is attained if the two assumptions made in deriving Stokes equation from Navier-Stokes equation, viz., $\lambda \to 0$ and $Re_\infty \to 0$ hold simultaneously. If the first limit holds, but the second one does not, deviation of $F_D/F_S$ from unity is merely a measure of inertial effect. In contrast, $V_t/V_\infty \to 1$ in the limit $\lambda \to 0$ irrespective of the value of $Re_\infty$. Its value depends on $Re_\infty$ only if $\lambda$ is not zero. These two ratios are related and are reciprocal of each other in the limit $Re_\infty \to 0$ (Eq. 8 and 10).

**1.7 Structure of the paper**: This paper is divided into several sections: (a) In section 2 we describe the experimental arrangements used for position measurement and terminal velocity calculation of a falling ball in a tube, (b) In sec. 3 the method of calculation of $\mu_\infty$ from measured values of $(V_t)$ is described. The core of the method is to calculate the correction factor that must be applied to the apparent Stokes viscosity coefficient. The method uses available velocity ratio data to calculate the less accessible viscosity ratio which is the desired correction factor. (c) In section 4 results obtained with our set-up on fluids of known $\mu_\infty$ are reported. These values are used to test the method of calculation of $\mu_\infty$ from $V_t$. The method is further tested with published high Reynolds number terminal velocity data on water. Problems specific to handling high Reynolds number data are assessed. Issues related to accuracy and precision are addressed. The method is shown to give high accuracy results well beyond the creepy flow regime to which falling ball viscometry is currently restricted.

**2. Experimental Section**

**2.1 Detection of ball position:** The position of the falling ball is measured by an optical method as well as by video photography. The optical method uses interruption by the falling ball of an infrared light beam that passes through the liquid. 8 optical detection stations each comprising of an infrared LED source and an aligned infrared detector are positioned at equidistant locations along the length of the tube that contains the liquid through which the ball falls. The interruptions are recorded through appropriate circuitry and a digital oscilloscope in a LabView environment in a PC. In this set up one can make measurements on transparent as well as turbid fluids. Off-axis ball movement cannot be detected.

Videos have been recorded 1080p (1980x1080) resolution at two frame rates viz., 30 and 250 FPS. The higher frame rate has been used in the cases where the total fall time is less that 1second. The axes of video (available in a camera in the form of grids) are aligned with those of tube to reliably record the verticality of falling ball trajectory in the recorded video. An incandescent tube light of length similar to that of the tube is placed behind the tube. The tube is wrapped with paper of appropriate thickness and colour, in order to diffuse light, maintain uniformity of light intensity throughout the tube and control its magnitude. With the whole tube uniformly illuminated, it is possible to study off-axis movement with ease.

**2.2 Mechanical details:** We have used three different diameter (internal) tubes viz. $40, 30$ and $16.5 mm$, each of length $70 cm$. Temperature of the fluid is controlled to $\pm 0.1°C$ by circulating water through an annular chamber around the tube. Room temperature was maintained close to the desired temperature. The ball was dropped in the liquid contained in the tube through precisely positioned holes in Perspex caps which are immersed in the liquid under test so that the ball does not encounter a medium discontinuity during its fall and to avoid formation of air bubbles. The ball was equilibrated in the same liquid before being dropped. A digital micrometer (least count $0.001\ mm$) was used for the measurement of ball radii. The non-sphericity is barely measurable with this least count. The manufacturer specified tolerance on degree of non sphericity is $0.25\ \mu m$. In accordance with this, the ball diameter is quoted to a precision of $1\mu m$. Tube diameter has been measured with a vernier caliper with least count of $0.001\ mm$. The tube radius was confirmed to be uniform by diameter measurement inside the tube and by letting a circular Perspex disc of radius equal to the tube radius at the mouth, fall all the way smoothly. The ratio $\lambda$ is calculated with the radius at the mouth as the tube radius. At other locations on the tube axis, a maximum nonuniformity is estimated to be $< 1\%$ on the lower side of the mouth radius. $\lambda$ is specified to third place of decimal. The verticality of the tube is achieved with use of spirit level and hanging bob, checked with the centerline fall of the smallest diameter ball terminating at the pre-marked center of a disc fitted at the bottom of the tube and further confirmed in more quantitative terms in the Video of the falling ball (Fig. 1).

In a separate paper we communicate the details of the measurement set-up and technical details of the two ball position detection methods.

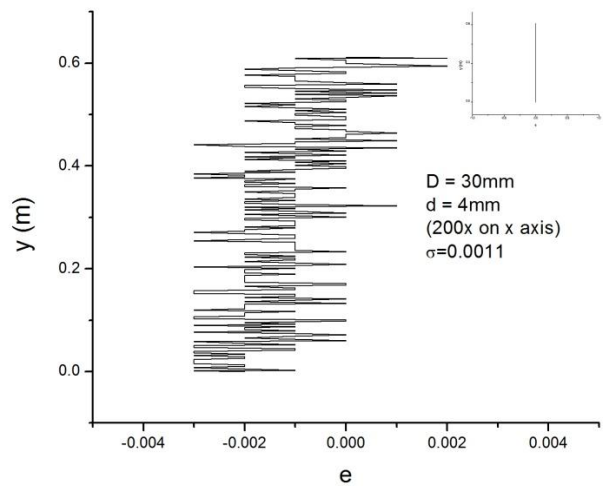

**Figure 1: Trajectory of 4mm ball in 30mm tube (from video); x axis is shown as 200x zoomed view. Inset shows full eccentricity view.**

**3. Method**

In the subsections that follow we discuss different aspects of the method we have developed and tested.

### 3.1. Relations between ratios

**3.1.1 $F_D/F_S$ and $\mu_S/\mu_\infty$:** The definition of $\mu_S$ (Eq.3) written in an equation structure is

$$F_D = 6\pi r \mu_S V_t \qquad (7)$$

where $V_t$ is measured in a narrow tube. This definition and Eq. 6 gives (Sutterby [38])

$$\frac{F_D}{F_S} = \frac{\mu_S}{\mu_\infty} \qquad (8)$$

an equality that holds at all $\lambda$ and $Re_\infty$.

**3.1.2 $\mu_S/\mu_\infty$ and $V_t/V_\infty$:** As already noted, for a given ball-fluid combination the sum of the force of gravity and that of buoyancy must equal frictional force in condition of steady fall. The sum is independent of whether the falling ball is in a narrow tube or in a boundless medium. One can then equate $F_D$ in a narrow tube (Eq. 7) to that in a boundless medium (substitution of $V_t = V_\infty$ in Eq. 6 obtains $(F_S)_{\lambda \to 0} = 6\pi\mu_\infty r V_\infty$)

$$6\pi\mu_S r V_t = 6\pi\mu_\infty r V_\infty \left(\frac{F_D}{F_S}\right)_{\lambda \to 0} \qquad (9)$$

It follows:

$$\frac{\mu_S}{\mu_\infty} = \left(\frac{V_t}{V_\infty}\right)^{-1} \left(\frac{F_D}{F_S}\right)_{\lambda \to 0} \qquad (10)$$

This is the desired relation. We recognize L.H.S. of Eq. 10 as $\left(\frac{F_D}{F_S}\right)_{\lambda \neq 0}$ (by Eq. 8) and conclude that the ratio of $\left(\frac{F_D}{F_S}\right)_{\lambda \neq 0}$ and $\left(\frac{F_D}{F_S}\right)_{\lambda \to 0}$ is $\left(\frac{V_t}{V_\infty}\right)^{-1}$, a relation we refer to in sec. 1.6.

We use abbreviations

$$\frac{\mu_\infty}{\mu_S} = f^\mu, \frac{V_t}{V_\infty} = f^V \qquad (11)$$

Then Eq. 10 becomes

$$f^V/f^\mu = \left(\frac{F_D}{F_S}\right)_{\lambda \to 0} \qquad (12)$$

Experimental information is available on the two factors that appear on the RHS of Eq. 10. The data have been parameterized (section 3.2) in terms of $\lambda$, a geometric parameter pertaining to the ball and the tube and $Re_\infty$, which is specific to a ball-liquid pair. It is therefore, possible to calculate (Eq. 10) $\mu_S/\mu_\infty$ over the whole range of $\lambda$ and a wide range of $Re_\infty$. Using $V_t$ one calculates only $Re_S$. We give below the method of calculating $Re_\infty$ from $Re_S$.

### 3.1.3 $Re_S$, $Re$ and $Re_\infty$ : Eqs. 2, 4, 5 and Eqs. 11, 12 give

$$Re_\infty = Re_S \left(\frac{V_\infty}{V_t}\right)\left(\frac{\mu_S}{\mu_\infty}\right) = Re_S/(f^V f^\mu) = \left(\frac{Re_S}{(f^V)^2}\right)\left(\frac{F_D}{F_S}\right)_{\lambda\to 0} \quad (13)$$

$$Re = Re_S(\mu_S/\mu_\infty) = Re_S/f^\mu \quad (14)$$

With $V_t$, only $Re_S$ can be calculated ; $Re$ and $Re_\infty$ are obtained from $Re_S$, with Eq. 13 and 14 if functional dependence of $f^V$ and $f^\mu$ on $Re_\infty$ (or $Re$) and $\lambda$ are known. The relation between $Re_\infty$ and $Re$ (Eq. 2 and 4) is given in Eq. 15;

$$Re_\infty = Re/f^V \quad (15)$$

With a known $\lambda$ and $Re_S$, substitution of the functional forms of $f^V(Re_\infty, \lambda)$ and $f^\mu(Re_\infty, \lambda)$ in Eq. 13, reduces it to an equation in a single unknown $Re_\infty$. Similarly substitution of the functional form of $f^\mu(Re, \lambda)$ in Eq. 14 reduces it to an equation in a single unknown $Re$.

In order to solve Eq.13, we recast the equation in the following form:

$$\left(\frac{Re_S}{(f^V(Re_\infty,\lambda))^2}\right) = \frac{Re_\infty}{\left(\frac{F_D}{F_S}\right)_{\lambda\to 0}} \quad (13a)$$

The values of $Re_\infty$ at the intersection points of the plots of the two functions of $Re_\infty$, given in RHS and LHS, of Eq. 13a, are roots of Eq. 13. Points of intersection of RHS and LHS of Eq.14 are roots of Eq. 14. $f^V, f^\mu$ are evaluated at the values of $Re_\infty, Re$ at the point(s) of intersection (and $\lambda$). We refer to these methods as 'graphical-methods'. The number of roots in both cases depend on the functional form of the expressions, determined by those of $f^V(Re_\infty, \lambda)$, $(F_D/F_S)_{\lambda\to 0}$ in Eq. 13 and $f^\mu(Re, \lambda)$ in Eq. 14. We give these functional forms in section 3.2 and consider the issue of number of roots in section 3.4.5.

For a given ball liquid pair, characterised by a $Re_\infty$ R.H.S. of Eq. 13a is $\lambda$ independent. The ratio on the LHS is tube diameter independent although the numerator and denominator are not. In the limit $\lambda \to 0$ ($f^V \to 1$) and $Re_S/(f^V)^2$ equals $Re_S$, the Stokes Reynolds number.

### 3.2 Correlation formulae:

#### 3.2.1 $(F_D/F_S)_{\lambda \to 0}$ and $Re_\infty$ : Clift equation and Cheng equation:

Clift et. al. [1] in a thorough summary of experimental results performed by different workers over many years recommend formulae for $(F_D/F_S)_{\lambda \to 0}$ over a wide range of Reynolds number with excellent predictive accuracy. Functional form of the formulae depend on $Re_\infty$ range. Subsequent research on the design of these formulae (Turton and Levenspiel [48], Brown and Lawler [49], Cheng [40]) focus on simplification of their structure to achieve range independence and incorporate small improvements in their predictive accuracy. We have used the equation of Cheng [40] given in Eq. 16, because of its range independent functional form ($Re_\infty < 2 \times 10^5$). The equation of Clift [1] valid in the range $Re_\infty = 0 - 20$, is given in Eq. 17.

$$\left(\frac{F_D}{F_S}\right)_{\lambda \to 0} = (1 + 0.27 \, Re_\infty)^{0.43} + \frac{Re_\infty}{24} 0.47[1 - \exp(-0.04 Re_\infty^{0.38})] \tag{16}$$

$$\left(\frac{F_D}{F_S}\right)_{\lambda \to 0} = 1 + 0.1315 \, Re_\infty^{0.82 - 0.05 \log_{10}(Re_\infty)} \text{ for } (0.5 < Re_\infty < 20) \tag{17}$$

The force ratio is a monotonically increasing function of $Re_\infty$. A plot of the slope of Eq. 16 as a function of $Re_\infty$ is given in Fig. 2. The slope rapidly decreases at low $Re_\infty$ to a minimum value $0.01544$ (at $Re_\infty = 1000$) and then rises monotonically to a limiting value $0.01999$.

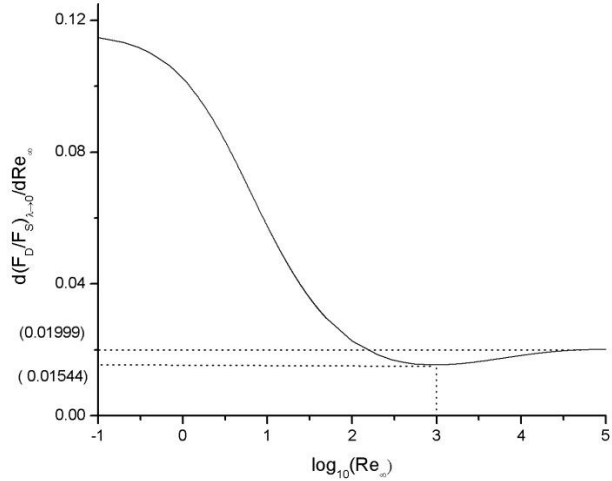

**Figure 2: Variation of slope of $(F_D/F_S)_{\lambda \to 0}$ with respect to $Re_\infty$ as a function of $Re_\infty$.**

**3.2.2 $f^V(Re_\infty, \lambda)$, DiFelice equation:** DiFelice [33] proposed an equation for wall factor $f^V(\lambda)$ that accomodates inertial effect with the aid of a single $Re_\infty$ dependent parameter.

The equation is:

$$f = \left(\frac{1-\lambda}{1-0.33\lambda}\right)^\alpha \tag{18a}$$

where $f^V = V_t/V_\infty$ and $\alpha$, a function of $Re_\infty$ that quantifies the inertial effect is defined as

$$\frac{3.3-\alpha}{\alpha-0.85} = 0.1 Re_\infty \tag{18b}$$

We observe that $\alpha$ is 0.85 in the limit of very large $Re_\infty$ and is 0.80 at $Re_\infty = 100$, i.e., it nearly saturates at $Re_\infty = 100$. This early saturation is not consistent with experimental results in the intermediate $Re_\infty$ regime. This deficiency was rectified in a subsequent paper of Kehlenbeck and DiFelice [34], henceforth referred to as K &D. The equation, henceforth called K-D equation, gives satisfactory results over a very wide $Re_\infty$ range (K & D, Chhabra [32]). It forms an important component of this paper. This improvement necessitated introduction of a second $Re_\infty$ dependent parameter in K-D equation.

**3.2.3 $f^V(Re_\infty, \lambda)$, K-D equation :** The two parameter K-D equation is

$$f_{KD}^V = \frac{(1-\lambda^p)}{\left(1+(\lambda/\lambda_0)^p\right)} \tag{19a}$$

where $\lambda_0$ and $p$, are defined as follows:

$$\frac{(\lambda_0 - 0.283)}{(1.2 - \lambda_0)} = 0.041 Re_\infty^{0.524} \tag{19b}$$

$$p = 1.44 + 0.5466 Re_\infty^{0.434} \quad (Re_\infty \leq 35) \tag{19c}$$

$$p = 2.3 + 37.3 Re_\infty^{-0.8686} \quad (Re_\infty \geq 35) \tag{19d}$$

These formulae summarize experimental data on $V_t/V_\infty$ in the parameter range $2 < Re_\infty < 185$, $0.1 < \lambda < 0.9$. However, K &D have shown that their equation holds at $Re_\infty$ as high as 10,000 and as low as 0.01 with nearly equal predictive accuracy. A detailed examination that we report later in the paper shows that the accuracy at low $Re_\infty (\sim 0.01)$ is not as satisfactory as it is at higher $Re_\infty (\sim 10000)$. Correlation formulae designed to predict $f^V$ at a given $(\lambda, Re_\infty)$ have been proposed earlier (Chhabra [32]), but that by K &D is of special interest to our work because it correlates $f^V$ over nearly whole $Re_\infty$ range. These authors determined $f^V$ in the intermediate $Re_\infty$ range in which available experimental data were scanty. Their parametrization of $f^V$ in this range is the only available formula of high predictive accuracy.

In contrast to DiFelice equation, K-D equation saturates at a much larger $Re_\infty$, viz. $10^5$ ($\lambda_0 = 1.2$ as $Re_\infty \to \infty$, 0.6 at $Re_\infty = 100$, ~ 1.15 at $Re_\infty = 10^5$). This feature makes K-D equation successful in the intermediate $Re_\infty$ range, where $f^V$ changes systematically, though slowly. The experimental data at different values of $Re_\infty$ and their agreement with the predictions of K-D equation are given in KD. In Fig. 3, we show plots of $f^V(\lambda)$, a decreasing function, at different fixed values of $Re_\infty$ as given by K-D equation. The slow saturation is apparent. It is an increasing function of $Re_\infty$, which saturates at large $Re_\infty$ (~$10^4$) for a fixed $\lambda$ (Fig. 4 of F&W). The $f^V(\lambda)$ plots in the intermediate range of $Re_\infty$ are S-shaped. At large $Re_\infty$ the shape is hyperbolic.

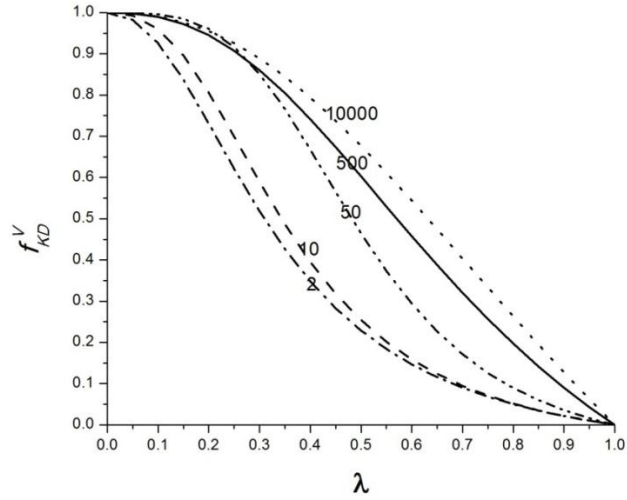

**Figure 3: Functional dependence of $f^V$ on $\lambda$ as given by K-D equation for a wide range of $Re_\infty$.**

**3.2.4 $f^\mu$ ($Re,\lambda$): Wham Equation:** Wham et. al. [41] use finite element method for simulation studies on a falling ball in a liquid medium contained in an infinitely long narrow tube ($0.08 < \lambda < 0.7$, $Re < 200$) open at both ends (no end effect), to obtain dependence of $F_D/F_S$ ( = $(f^\mu)^{-1}$), Eq. 8 and 11) on $\lambda$ and $Re$. The governing equations are Navier-Stokes equation and the continuity equation. Their data is represented by Eq. 20 and henceforth is referred to as Wham equation.

$$\frac{F_D}{F_S} = \left(1 + 0.03708 * \left(\frac{Re}{2}\right)^{(1.514 - 0.1016 \ln \frac{Re}{2})}\right) * \left(\frac{(1 - 0.75857\lambda^5)}{(1 - K*\lambda + 2.0865*\lambda^3 - 1.7068*\lambda^5 + 0.72603*\lambda^6)}\right) \quad (20a)$$

where, $K = 0.6628 + 1.458 * \exp\left(-0.05635 * \frac{Re}{2}\right)$ \quad (20b)

The corresponding expression in Wham et. al. [41] uses a definition of Re ($= \rho V_t d/\mu_\infty$, where r is ball radius) that differs from the usual definition, we give in Eq. 4. In Eq. 20 we use $Re$ as defined in Eq. 4. L.H.S. of Eq. 20a is $(f^\mu)^{-1}$. By Eq. 12 $f^V$ equals the ratio of Eq. 16 and 20a.

### 3.3 The Limiting forms:

**3.3.1 Cheng Equations**: In the limit $Re_\infty \to 0$ R.H.S. of Eq. 16 approaches unity. In the limit of very large $Re_\infty$ ($\sim 10^5$) its approximate form is: $0.5\sqrt{Re_\infty} + \frac{Re_\infty}{51}$.

**3.3.2 Wham Equation:** The $Re \to 0$ limit of Wham equation (Eq. 20) is

$$\left(\frac{F_D}{F_S}\right)^W_{Re\to 0} = \frac{(1-0.75857\lambda^5)}{(1-2.1208\lambda+2.0865\lambda^3-1.7068\lambda^5+0.72603\lambda^6)} \tag{21}$$

The H-S equation is

$$\left(\frac{F_D}{F_S}\right)_{Re\to 0} = \frac{(1-0.75857\lambda^5)}{(1-2.1050\lambda+2.065\lambda^3-1.70678\lambda^5+0.72603\lambda^6)} \tag{22}$$

No assumption is implicit in obtaining this limit $Re \to 0$. Eq. 21 has an identical structure and nearly equal coefficient values (the coefficient linear in $\lambda$ is slightly different, $-2.105$ instead of $-2.1208$ in Eq. 21) as the H-S equation (Eq. 22). Its range of validity is $\lambda < 0.8$ and $Re < 2$. The limit $Re \to 0$ is the same as the limit $Re_\infty \to 0$. The function used in fitting simulation results in the work of Wham et. al. [41] appears to have been chosen in such a way that in the limit $Re_\infty \to 0$ it 'nearly' reduces to H-S equation.

An expression for $(F_D/F_S)_{Re\to 0}$ can be derived from Wham equation if we (incorrectly) assume that Eq. 20 holds in the limit $\lambda \to 0$. Then the second bracketed term of Eq. 20a is unity and $Re$ in the first term is replaced by $Re_\infty$. This equation gives values that differ considerably from those given by Eq. 16 to show that Eq. 20a in the limit $\lambda \to 0$ is not valid.

**3.3.3 DiFelice equation and K.D. equation:** In the limit $Re_\infty \to 0$ DiFelice equation assumes a limiting form with $\alpha = 3.3$ (Eq. 18a). K-D equation assumes a limiting form with $\lambda_0 = 0.283$ (Eq. 19b), $p = 1.44$ (Eq. 19c). In this limit $f^V$ equals $f^\mu$.

**3.3.4 $f^V$ and $f^\mu$ at large $Re_\infty$:** With increase in $Re_\infty$, $f^V_{KD}$ is progressively insensitive to increase in $Re_\infty$ (Fig. 4 of F&W and Fig. 1 of DiFelice [33]). $Re_\infty$ independence of $f^V(\lambda)$ at large $Re_\infty$ has been noted (Clift [1], Munroe [50], Arsenijevic [51]) and has a functional form different from that in the $Re_\infty \to 0$ limit. The large $Re_\infty$ limit is $\lambda$ dependent ($Re$ of 53 at $\lambda \sim 0.05$, $Re > 10^4$ at $\lambda \sim 0.8$). Although $f^V$ is $Re_\infty$ insensitive, $f^\mu$ is not, since $(F_D/F_S)_{\lambda \to 0}$ increases with $Re_\infty$. Wham equation is not valid beyond $Re = 200$. Wham's equation of $f^\mu$ remains $Re_\infty$ dependent upto its upper limit of validity, $Re = 200$.

**3.3.5 Stokes limit:** We deduce some features of approach to Stokes limit with the aid of equations cited above We note (i) for $\lambda \ne 0$, $f^V = (V_t/V_\infty) < 1$ (Eq. 18), for any $Re_\infty$ (ii) for $Re_\infty \ne 0$, $(F_D/F_S)_{\lambda \to 0} > 1$ (Eq. 16, 17), we use Eq. 10 and $f^V < 1$ shown in (i) to conclude, that for any $\lambda$ and $Re_\infty$, $(F_D/F_S)_{\lambda \ne 0} > 1$, $f^\mu < 1$ (iii) in the limit $Re_\infty \to 0$, $(F_D/F_S)_{\lambda \to 0} = 1$ (Eq. 16, 17), as a result $(F_D/F_S) = (V_t/V_\infty)^{-1}$ for all $\lambda$; an equality referred to earlier (iv) $\lambda \to 0$ is a sufficient condition for $(V_t/V_\infty) \to 1$ (Eq. 18); $Re_\infty \to 0$ is not necessary (v) combining (iii) and (iv) we conclude that $(F_D/F_S) \to 1$ only if conditions for validity of Stokes equation, $\lambda \to 0$ and $Re_\infty \to 0$ hold simultaneously (vi) Stokes limit$(F_D/F_S) \to 1$, therefore is not implied by $(V_t/V_\infty) \to 1$, $Re_\infty \to 0$ must also hold.

**3.4 Calculation Schemes:** In this section, we detail procedures of calculation that are used in later developments. The same quantity is calculated by several different procedures and their accuracy is assessed.

**3.4.1 $\mu_\infty \to Re_\infty, f^V, f^\mu, V_t$: Cheng equation and K-D (or Wham) equation:** We use a known value of $\mu_\infty$ to calculate $Re_\infty$ as outlined below. We use (i) Eq. 16 (ii) the knowledge that $F_D$ in state of steady fall equals the sum of the forces of gravity and buoyancy (iii) Eq. 6 with substitution of $V_\infty$ in place of $V_t$ to obtain $(F_S)_{\lambda \to 0}$ and (iv) Eq. 2 which defines $Re_\infty$ in terms of $V_\infty$. We then obtain an equation in a single unknown variable $Re_\infty$, which is solved iteratively to obtain numerical value of $Re_\infty$ (and then $V_\infty$ from Eq. 2). This calculation requires as input the accurately known values of physical parameters of the ball-liquid pairs, namely $\mu_\infty, d, \rho_b$ and $\rho_l$. Using this value of $Re_\infty$ (designated $Re_\infty^\mu$) and that of $\lambda$ we calculate $f_{KD}^V$ by Eq. 19. Use of the value of $Re_\infty^\mu$, Eq. 16, Eq. 12 and the value of $f_{KD}^V$ gives $f_{KDC}^\mu$. These values are designated $f_{KD}^V(\mu)$, $f_{KDC}^\mu(\mu)$ respectively, to distinguish them from their counterparts defined in sec. 3.4.2 and 3.4.3. One substitutes the value of $Re_\infty$ in Eq. 2 to get $V_\infty$ (designated $V_\infty^\mu$) and further Eq. 11 and the value of $f^V$ to calculate $V_t$.

If Wham equation is being used, we must still use Cheng equation to calculate $Re_\infty$ from $\mu_\infty$. We then solve Eq. 15 iteratively, as detailed below, to calculate numerical value of $Re$ and $f^V$ (referred to as $f_{WC}^V$) given the value of $Re_\infty$. The functional form of $f_{WC}^V$ is given by the product of RHS of Eq. 16 and Eq. 20a. The expression is reduced to a function of a single variable $Re$ by substitution of the numerical value of $Re_\infty^\mu$, as calculated from Eq. 16, and that of $\lambda$. $Re$ is then calculated. Eq. 4 gives value of $V_t$ from that of $Re$. Value of $Re$ and that of $\lambda$ with Eq. 20 gives the value of $f_W^\mu$.

**3.4.2 $V_t \to Re_\infty, f^V, f^\mu, \mu_\infty$: K-D equation and Cheng Equation:** We substitute the forms of Eq. 16 and Eq. 19 in Eq. 13 which is restructured in Eq. 13a. $Re_S$ is an experimentally determined input to this equation. The solutions are $Re_\infty$, $f^V$, $f^\mu$. This combination of equations henceforth referred to as KDC equation, is solved most transparently by a graphical method given in sec. 3.1.3 (also Fig. 4). One scans $Re_\infty$ over the whole range $(0 - 10^5)$, obtain the intersection points of the plots of RHS and LHS of Eq. 13a. With $Re_\infty$ of each intersection point and $\lambda$ one calculates the respective $f^V$ and $f^\mu$. With use of experimentally determined $\mu_S$, one obtains $\mu_\infty$ for each $Re_\infty$. Choice between the roots, if there is more than one, is based on additional information (sec. 3.4.5). In essence knowledge of $V_t$, Eq. 16 and 19 enables us to calculate $V_\infty$ without performing a velocity measurement in infinite medium.

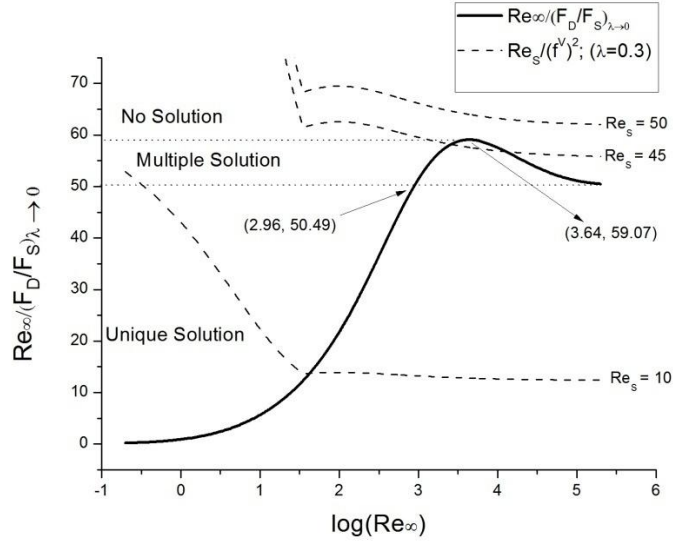

**Figure 4: Plot of $Re_\infty/(F_D/F_S)_{\lambda\to 0}$ as a function of $Re_\infty$; three possible cases.**

In an equivalent procedure we may adapt the method of calculation of $V_t$ from $\mu_\infty$ (sec. 3.4.1), in reverse to calculate $\mu_\infty$ from $V_t$. One (i) scans $\mu_\infty$ over a range; (ii) calculates $V_t$ for each $\mu_\infty$; then (iii) correct value of $\mu_\infty$ is one that gives experimental value of $V_t$. These two methods are convenient in search of multiple solutions of Eq. 13a by KDC procedure. A standard iterative procedure can also be used to determine $Re_\infty$, $f^V_{KD}$ and $f^\mu_{KDC}$ together in a single step. $\mu_S$ and $f^\mu$ give $\mu_\infty$ by Eq. 11.

$Re_\infty$, $f^V$, and $f^\mu$ determined from $V_t$ and K-D equation are designated $Re_\infty^{V_t}(KD)$, $f^V_{KD}(V_t)$ and $f^\mu_{KDC}(V_t)$ respectively.

**3.4.3 $V_t \to f^\mu, \mu_\infty, f^V, Re_\infty$: Wham equation:** We substitute in Eq. 14, the functional form of $f^\mu_W$ specified in Eq. 20 (Wham equation), solve it to calculate, for a given $V_t$, $Re_s$ the numerical value of $Re$ and $f^\mu_W$ and thence $\mu_\infty$. In this procedure we do not calculate $f^V_{WC}$ explicitly in the process of calculating $\mu_\infty$.

In order to calculate values of $Re_\infty$ and $f_{WC}^V$ we use the route specified below. The values are obtained together in a single step as solution of Eq. 15. The expression of $f_{WC}^V$ we use, is product of RHS of Eq. 16 function of $Re_\infty$ and that of Eq. 20, a function of ($Re$, $\lambda$). The expression is reduced to a function of a single unknown $Re_\infty$ by substitution of numerical values of $Re$ as calculated above and $\lambda$. We obtain values of $Re_\infty$ and $f_{WC}^V$ at a specific ($\lambda$, $Re_\infty$). This value of $Re_\infty$ is determined for $V_t$ data and Wham equation and is designated as $Re_\infty^{V_t}$ (Wham).

**3.4.4 $f_{WC}^V$ :** In sec. 3.4.1 we calculate $f_{WC}^V$ given $Re_\infty$ as an input. In sec. 3.4.3 we calculate $f_{WC}^V$ given $Re_S$ (or $V_t$) as an input. They are designated $f_{WC}^V(\mu)$ and $f_{WC}^V(V_t)$ respectively.

**3.4.5 Uniqueness of calculated parameters:** We consider the graphical method of solution of Eq. 13, using equation structure of Eq. 13a, as given in sec. 3.4.2. The roots of Eq. 13a are obtained from intersection points of $(Re_\infty/(F_D/F_S)_{\lambda \to 0})$, and $Re_S/(f^V)^2$ (RHS and LHS of Eq. 13a as functions of $Re_\infty$).

The function $Re_\infty/(F_D/F_S)_{\lambda \to 0}$, RHS of Eq. 13a, shown as a plot in Fig. 4 has a single maximum with a function value 59.07 for $Re_\infty = 4363$. The values fall monotonically on both sides of the maximum; to zero as $Re_\infty \to 0$ and to $\sim 50.5$ as $Re_\infty$ assumes large values $\sim 10^5$, the upper limit of validity of Eq. 16.

A plot of $Re_S/(f^V)^2$ vs. $Re_\infty$ is determined by the experimental value of $Re_S$, a fixed input and has the shape of $1/(f^V)^2$ at a fixed $\lambda$ as a function of $Re_\infty$. For given values of physical parameters of the ball-liquid pair $\rho_b$, $\rho_l$ and $d$, $Re_S$ assumes larger values for liquids of low $\mu_\infty$ and $f^V$ is increasingly smaller as $\lambda$ increases for a given $Re_\infty$. The plot of $Re_S/(f^V)^2$ vs. $Re_\infty$ shifts upwards to larger values on y-axis with increase in $\mu_\infty$ and/or increase in $\lambda$. Fidleris and Whitmore [36] give the shape of $1/f^V$ as a function of $log(Re)$ for several values of $\lambda$ in Fig. 3 of their paper. The plots are flat at small values of $log(Re)$ and decrease to show another flat region (which is not as flat) at high values of $log(Re)$. The flat region at low $log(Re)$ extends up to larger values of $log(Re)$ as $\lambda$ decreases. The difference in the values of $1/f^V$ in the two flat regions increases with increasing $\lambda$. Thus, at smaller $\lambda$ ($\leq 0.2$), $1/(f^V)$ vs. $Re$ has a very flat appearance over the whole $Re$ range. At a larger $\lambda$ its decrease from one flat region to the other and consequent saturation becomes increasingly more apparent. In this larger $\lambda$ family the value of $Re$ at which saturation is observed increases with increasing $\lambda$. The shape of $1/(f^V)^2$ vs. $log(Re_\infty)$ will be similar. In many $Re_\infty$ scans the low $Re_\infty$ flat region is omitted (being $\ll Re_\infty$ at first intersection) and one observes an initial decrease followed by a somewhat flat region on $1/(f^V)^2$ plot (Fig. 4).

An examination of Fig. 4 shows that multiple solutions are obtained only for large values of $Re_\infty$ (x-coordinate) and $Re_S/(f^V)^2$ (y-coordinate); $Re_\infty > \sim 907$, $50.5 \leq Re_S/(f^V)^2 \leq 59.07$ for the first intersection point. In this range of values scanning beyond the first point of intersection keeps $Re_S/(f^V)^2$ plot a straight line parallel to $Re_\infty$ axis, followed by a second point of intersection at a larger value of $Re_\infty$. Beyond this second point the plot of $Re_S/(f^V)^2$ remains parallel to the asymptotic straight line plot of $Re_\infty/(F_D/F_S)_{\lambda \to 0}$. As a consequence, more than two alternate values of $Re_\infty$ will never be obtained as solutions of Eq. 13a.

At the intersection point the equality specified by Eq.13a holds and $Re_S/(f^V)^2$ is determined by $Re_\infty$ alone, i.e., by the ball-liquid pair and is independent of the tube diameter. One can however choose a value of $Re_\infty$ by identifying a suitable ball radius and density which for a given liquid will give an unique value or a pair of values of $Re_\infty$ as solution of Eq. 13a.

The two roots merge if $Re_S/(f^V)^2 = 59.07$ at $Re_\infty = 4363$. The two plots touch each other at (4363, 59.07) and do not cross again. With even a small extent of error-contamination, already large values of $Re_S/(f^V)^2$ at a large $\lambda$ may exceed 59.07 at $Re_\infty = 4363$. The plot of $Re_S/(f^V)^2$ in such a case remains above that of $Re_\infty/(F_D/F_S)_{\lambda \to 0}$ for all values of $Re_\infty$. There exists no point of intersection and then Eq.13a has no solution (Fig.4).

The two values of $Re_\infty$ that correspond to the two intersection points, for a given $\lambda$ give two different values of $f^V$. The feature of two roots arises only if the value of $Re_\infty$ at the first point of intersection is large ($Re_\infty > 907$), as is the case with low viscosity liquids. $f^V$ is insensitive to $Re_\infty$ for large values of $Re_\infty$. As a result the two values of $Re_\infty$ correspond to closely spaced $f^V$. Irrespective of whether the two values of $f^V$ are closely spaced or are significantly separated, the corresponding values of $f^\mu$ will be different because of significantly unequal $Re_\infty$. The two calculated values of $\mu_\infty$ are then different. The larger value of $Re_\infty$ obtained at the second point of intersection will result in a smaller value of $\mu_\infty$. In the cases we study and discuss in sec. 4.3, their values are such that an unequivocal choice is possible. The separation in values between the two alternate $Re_\infty$ and corresponding $\mu_\infty$ decrease with increase in $Re_S$ and $\lambda$. If the difference is small, the choice may not be so unequivocal (Fig.4). Choice of a smaller ball diameter, and a smaller ball density (Teflon ball instead of Steel balls) can take a system out of the multiple root regime into the unique root regime or from two closely spaced roots to two separated roots. The ambiguities are then satisfactorily resolved.

**3.4.6 $(\mu_\infty, V_t) \to (f^V, f^\mu)$:** The estimates of $f^V$, $f^\mu$ made in sec. 3.4.1, 3.4.2, 3.4.3 use as experimental input only $\mu_\infty$ or only $V_t$. One may use both $V_t$ and $\mu_\infty$ to calculate them. $f^\mu$ has been calculated as ratio of $\mu_\infty$ (determined experimentally using independent viscometers) and $\mu_S$, that uses $V_t$ (Sutterby [38]). $f^V$ is calculated as follows: $V_\infty$ is calculated from already 'known' $\mu_\infty$ and Eq. 16 (sec. 3.4.1). With Eq. 11 and experimental $V_t$ one obtains $f^V$. The ratios $f^V$, $f^\mu$ so calculated are designated $f^V(\mu, V_t)$ and $f^\mu(\mu, V_t)$ respectively. The $(f^V, f^\mu)$ pair are related by Eq. 12 and 16, with $Re_\infty^\mu$ calculated as in sec. 3.4.1 from 'known' values of $\mu_\infty$.

**3.4.7 Analysis of Error:** In 'Results and Discussions' we infer that the functional form of Eq. 16, 19, 20 used in calculations may require modification of their form to remove inconsistencies in calculated values. Of these three, Eq. 19 and 20 have a larger probability of having an inexact form than does Eq. 16 (sec. 4.3.1, 4.3.2). Scatter of accurate experimental values are present around those given by the 'best' of such functional forms. Inaccuracy so introduced is inherent in the choice of a functional form. In sec. 4.3.1 and 4.3.2 we conclude that the functional form of Eq. 16 be taken to be 'exact' but Eq. 19 or 20 may be amenable to small modifications. The method used for the analysis of this error transmission given in the following subsections, makes this assumption.

**3.4.7.1 $f^V$:** The values of $f^V(\mu, V_t)$ are not contaminated by errors in 'inexact' functional forms of Eq. 19 and 20. In addition to values of $\mu_\infty$, $V_t$, one uses only Eq. 16, which is a high accuracy representation of experimental data of $V_\infty$ (sec. 4.3.1). One can say that they are obtained from accurate experimental data alone.

We define the error in $f^V_{KD}$ as

$$\Delta f^V_{KD} = f^V(\mu, V_t) - f^V_{KD}(V_t) \tag{23a}$$

$\Delta f^V_{KD}$ is further decomposed into two components as defined below

$$\Delta f^V_{KD} = \Delta f^V_1 + \Delta f^V_2 \tag{23b}$$

$$\Delta f^V_1 = f^V_{KD}(\mu) - f^V_{KD}(V_t) \tag{23c}$$

$$\Delta f^V_2 = f^V(\mu, V_t) - f^V_{KD}(\mu) \tag{23d}$$

Corresponding error terms in Wham calculations are defined identically. Of the three $f^V$ terms that appear in Eq. 23c and 23d, $f^V_{KD}(\mu)$ and $f^V_{KD}(V_t)$ derive error form an inexact functional form of Eq. 19. $f^V_{KD}(\mu)$ uses Eq. 19 with $Re^\mu_\infty$ calculated from Eq. 16 alone (Eq. 19 not used). $f^V_{KD}(V_t)$ used Eq. 19 with $Re^{V_t}_\infty$, whose calculation further uses Eq. 19 (in Eq. 13a). As a result $Re^\mu_\infty \neq Re^{V_t}_\infty$ ($Re^{V_t}_\infty$ is less accurate) and $f^V_{KD}(\mu) \neq f^V_{KD}(V_t)$. $f^V(\mu, V_t)$, the correct value of $f^V$ can be thought of as having been obtained by substitution of $Re^\mu_\infty$ in the correct functional form of $f^V(Re_\infty, \lambda)$ (which may not be known) and $f^V_{KD}$ uses the same $Re^\mu_\infty$ in the incorrect form given in Eq.19. $\Delta f^V_2$ is an estimate of what may be called a purely form error.

**3.4.7.2 $f^\mu$:** In view of excellent accuracy of the functional form of Cheng formula (Eq. 16) for $(F_D/F_S)_{\lambda \to 0}$ (sec. 4.3.1) error in $f^\mu$ ($\Delta f^\mu$) is derived almost entirely for that of the functional form of $f^V(\Delta f^V)$. Translation of $\Delta f^V$ to $\Delta f^\mu$ is executed in two parts. For K-D equation this is denoted as: $\Delta f^V_1 (KD) \to \Delta f^\mu_1 (KDC)$, $\Delta f^V_2 (KD) \to \Delta f^\mu_2 (KDC)$. The transformations are dependent on the values of different estimates of $Re_\infty$. The relevent expressions are given below

$$\Delta f^\mu(KDC) = \Delta f^\mu_1(KDC) + \Delta f^\mu_2(KDC) \tag{24a}$$

$$\Delta f_1^\mu(KDC) = \left(\frac{f_{KD}^V}{(F_D/F_S)_{\lambda\to 0}}\right)_{Re_\infty=Re_\infty^\mu} - \left(\frac{f_{KD}^V}{(F_D/F_S)_{\lambda\to 0}}\right)_{Re_\infty=Re_\infty^{V_t}} \quad (24b)$$

An inexact form of $f_{KD}^V$ (Eq. 19) results in $Re_\infty^{V_t} \neq Re_\infty^\mu$. As a consequence the value of $f_{KD}^V$ and that of $(F_D/F_S)_{\lambda\to 0}$ change. These changes combine according to Eq. 24(b) to give $\Delta f_1^\mu(KDC)$

$$\Delta f_2^\mu(KDC) = f^\mu(\mu, V_t) - \left(\frac{f_{KD}^V}{(F_D/F_S)_{\lambda\to 0}}\right)_{Re_\infty=Re_\infty^\mu} \quad (24c)$$

We recognize the equality

$$f^\mu(\mu, V_t) = f^V(\mu, V_t)/(F_D/F_S)_{\lambda\to 0}^{Re_\infty^\mu} \quad (24d)$$

Eq. 24c then gives

$$\Delta f_2^\mu(KDC) = \Delta f_2^V(KD)/(F_D/F_S)_{\lambda\to 0}^{Re_\infty^\mu} \quad (24e)$$

Eq. 24d and 24e give

$$\Delta f_2^\mu(KDC)/f^\mu(\mu, V_t) = \Delta f_2^V(KD)/f^V(\mu, V_t) \quad (24f)$$

The ratio $\Delta f_2^\mu/\Delta f_2^V$ ($= 1/(F_D/F_S)_{\lambda\to 0}$) is always $< 1$ and becomes smaller as $Re_\infty$ increases.

The expression of $\Delta f_1^\mu$ (Eq. 24b) can be rewritten as

$$\Delta f_1^\mu/f^\mu = \left[1 - \frac{1-(\Delta f_1^V/f^V)}{1-\left(\Delta(F_D/F_S)_{\lambda\to 0}/(F_D/F_S)_{\lambda\to 0}^{Re_\infty^\mu}\right)}\right] \quad (25)$$

where, $\Delta(F_D/F_S)_{\lambda\to 0} = (F_D/F_S)_{\lambda\to 0}^{Re_\infty^\mu} - (F_D/F_S)_{\lambda\to 0}^{V_t}$;

$(\Delta f_1^V/f^V)$ is small in numerical value, more so at a large $Re_\infty$. The inequality $\left|\Delta(F_D/F_S)_{\lambda\to 0}/(F_D/F_S)_{\lambda\to 0}^{Re_\infty^\mu}\right| \ll 1$ holds for small $Re_\infty$, particularly at a smaller $\lambda$. One then obtains $\Delta f_1^\mu/f^\mu = (\Delta f_1^V/f^V) - \left(\Delta(F_D/F_S)_{\lambda\to 0}/(F_D/F_S)_{\lambda\to 0}^{Re_\infty^\mu}\right)$ i.e., $(\Delta f_1^\mu/f^\mu)$ is necessarily small and can have either relative sign. At a larger $Re_\infty$, $\left|\Delta(F_D/F_S)_{\lambda\to 0}/(F_D/F_S)_{\lambda\to 0}^{Re_\infty^\mu}\right|$ increases in magnitude remaining below unity and

$(\Delta f_1^V/f^V) \ll 1$. Irrespective of the sign of $\Delta(F_D/F_S)_{\lambda\to 0}/(F_D/F_S)_{\lambda\to 0}^{Re_\infty^\mu}$. One then obtains a relative amplification of $\Delta f_1^\mu/f^\mu$ with respect to $\Delta f_1^V/f^V$; there is a change of relative sign if it is positive and no change, otherwise.

A certain value of $\Delta f^V$ in two different $Re_\infty$ regimes (viz., $10^4$ as against 10) can result into values of $\Delta f^\mu$ of different orders. We observe these features in our analysis of data in lower $Re_\infty$ regimes in studies of Glycerol and Silicone Oil and that in larger $Re_\infty$ regime in study of Water. The magnitude of $\Delta f^\mu$ determines accuracy of $\mu_\infty$ determination. The above considerations made with KDC equation have their counterparts in calculations with Wham equation.

**4. Results and Discussions**

**4.1 Dependence of $Re_\infty$ on system parameters:** Although, the relationship between $f^V$ and $Re_\infty$ is known to high accuracy in the limit $Re_\infty \to 0$ and to a lesser accuracy level at finite $Re_\infty$, it may not be possible to perform experiments in this limit with an arbitrary liquid. There are restrictions on ball diameter imposed by difficulty of ball detection in a given technique and on density of ball material which must exceed liquid density. For a given liquid ($\mu_\infty$ and $\rho_l$) $Re_\infty$ can be written as a function of $\Delta\rho (= \rho_b - \rho_l)$ and $d$ using Eq. 9 as follows:

$$(g\,\Delta\rho/18)d^3 = (\mu_\infty^2/\rho_l)(Re_\infty\,(F_D/F_S)_{\lambda\to 0}) \qquad (26)$$

Given the physical parameters of the ball and the test liquid one can use Eq. 26 and Eq.16 to compute $Re_\infty$ from the physical parameters of ball and liquid,. One finds that $Re_\infty$ increases with $d$, $\Delta\rho$ (for a fixed $\rho_l$), $\rho_l$ (for a fixed $\Delta\rho$) and decreases with $\mu_\infty$ ( Figs. 5a, 5b).The increase of $Re_\infty$ with increasing $d$ is more rapid than quadratic. $Re_\infty/d$ is an increasing function of $d$. We use this result in sec. 4.2.2.

In Fig. 5a we note that with small $d(= 1mm)$, and $\Delta\rho = (5\,kgm^{-3})$ a virtual lower limit on grounds of ability to detect the ball position and availability of suitable ball material, one obtains a $Re_\infty \sim 2$ which rises to $Re_\infty \sim 50$ for $d = 4mm$ in a low viscosity liquid like water. The value of $Re_\infty$ is larger as $\Delta\rho$ for a given $d$ or $d$ for a given $\Delta\rho$ is increased. These problems are less acute for more viscous liquids. We infer that one must validate methods for calculation of $\mu_\infty$ by falling ball viscometry at values of $Re_\infty$ well beyond the 'creepy flow' regime, an exercise we have undertaken in this paper.

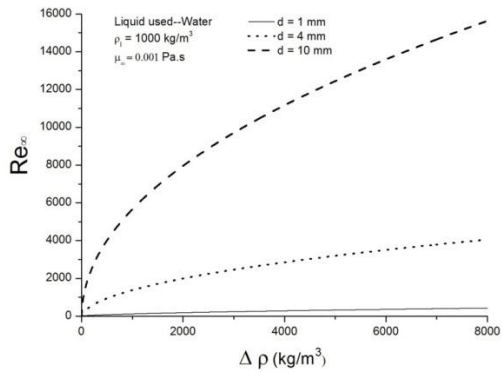
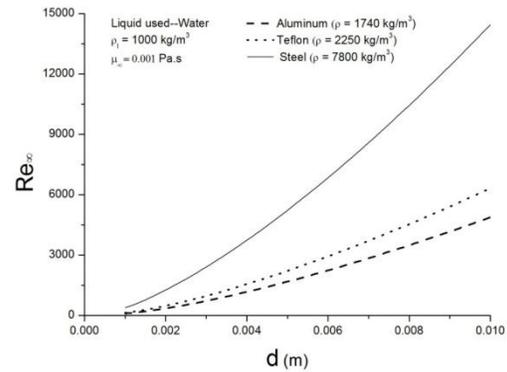

**Figure 5a:** $Re_\infty$ vs. $\Delta\rho$, difference in density between ball and liquid

**Figure 5b:** $Re_\infty$ vs. $d$, diameter of the falling ball

### 4.2 Our experiments at Lower $Re_\infty$:

**4.2.1 Terminal velocity data:** A combination of seven balls diameters ($d$) and three tube diameters ($D = 40, 30, 16.5\ mm$) are used in measurements of $V_t$. The values of $\lambda$ for different ball-tube combinations are given in Table 1. In Table 2 we summarize experimental results on $V_t$. Silicone oil and Glycerol at 30°C are the test liquids in two experiment sets. With Silicone oil, we vary $d$ for a fixed $D$ and also $D$ for a fixed $d$. With Glycerol, we carry out only the first variation in the tube with the largest $D$ ($= 40mm$). Only five values of $d$ ($\leq 14mm$) could be used in experiments with Silicone oil in the narrowest tube ($D = 16.5mm$). Reported experiments are performed with steel balls. Five measurements are performed for each ball-tube pair. The range of $\sigma$ in velocities in percent for Silicone oil in 40mm tube is $(0.01\%, 0.11\%)$; in $30mm$ tube is $(0.01\%, 0.11\%)$ and in $16.5mm$ tube is $(0.02\%, 0.19\%)$ respectively. The errors are independent of $D$. In measurements on Glycerol in 40mm tube the range is $(0.04\%, 0.94\%)$. In the case of Glycerol the error increases with increasing $\lambda$ and in all cases is larger than that for Silicone oil, still remaining small compared to most literature reports. The difference may arise from the more complex liquid structure of Glycerol. The typical values of $\sigma$ for ball density and ball diameter are 0.19% and ~0.01% respectively. These errors contribute negligibly to random errors of $V_t$, so does the uncertainty in liquid density (0.01%). The random error in estimating tube diameter is also of negligible consequence. The tube diameter non-uniformity which may contribute to a systematic error has been estimated to be less than 0.5% and does not alter our conclusions. Some literature reports on repeatability of $V_t$ measurements are: K-D, 5% (stop-watch); Brizard [7], 0.1% and less (CCD camera); Tran-Son-Tay et. al. [28], 2% in viscosity (ultrasonic); Flude and Daborn [52], 0.07%, (and 0.3% in viscosity) (Laser Doppler). Our precision is comparable to the very best of the reported value. In two other reports precision is not cited; values of accuracy quoted are: F&W, 0.3% (induction coil) and Subbaraman et. al. [54], 2%, (cathetometer and stop watch).

**Table 1: Values of λ for different ball and tube diameters**

| d (mm) | λ D=40mm | λ D=30mm | λ D=16.5mm |
|---|---|---|---|
| 4.000 | 0.100 | 0.133 | 0.242 |
| 5.000 | 0.125 | 0.167 | 0.303 |
| 7.000 | 0.175 | 0.233 | 0.424 |
| 9.000 | 0.225 | 0.300 | 0.545 |
| 11.000 | 0.275 | 0.367 | 0.667 |
| 14.280 | 0.357 | 0.476 | -- |
| 17.450 | 0.436 | 0.582 | -- |
| 20.620 | 0.516 | 0.687 | -- |

The $V_t$ values are precise to fourth decimal place and with $\sigma <$ 0.1% to third decimal place in experiments if $\sigma$ exceeds 0.1%. All values are quoted to third decimal place.

The values of $V_t$ in Table 2 show an increase in $V_t$ followed by a decrease. The same pattern has been reported by F&W in Fig. 1 of their paper for a low viscosity, high $Re$ liquid, namely water. The value of $d$ at which the maximum $V_t$ is observed (in a series of increasing $d$ at a fixed $D$) decreases with decreasing $D$. The $\lambda$ values at which maximum $V_t$ is observed in each of the tubes with $D = 40, 30$ and $16.5 mm$ with Silicone oil (and for $D = 40\ mm$ with Glycerol) are however, all very close ~0.5. With Silicone oil, the values of $V_t$ for the lower end $d = 5mm$ are comparable ($0.072, 0.065$ and $0.041 m/s$) for all three values of $D$. The increase to the maximum is the least in the narrowest tube, a maximum of $0.051 m/s$ is reached at $d = 7mm$, in comparison to a much larger maximum value of $V_t$ $(0.272, 0.156\ m/s)$ attained at a larger value of $d$ $(17.45\ and\ 14.28 mm)$ for tubes of larger $D$ $(40\ and\ 30 mm)$ respectively. The low values of $V_t$ attained for the largest value of $d$ in tubes of different $D$ $(20.62 mm$ in $D = 40, 30 mm$ and $11 mm$ for $D = 16.5 mm)$ differ from each other more significantly $(0.250, 0.077, 0.028)$ than their counterparts at the lower end $d$ $(5mm)$. The $\lambda$ values at the larger end $d$ differ more significantly than those in the lower end $d$. With a given $d$, as $D$ increases the value of $V_t$ shows a decrease for Silicone oil. The less viscous Glycerol shows a larger $V_t$ for a given $d$ and $D$ $(= 40\ mm)$.

**Table 2: Values of $V_t$, $\mu_s$, Res (low $Re_\infty$ systems)**

| d (mm) | Silicone Oil | | | | | | | | | Glycerol | | |
| | D= 40 mm | | | D = 30 mm | | | D = 16.5 mm | | | D= 40 mm | | |
| | $V_t$ | $\mu_S$ | $Re_S$ | $V_t$ | $\mu_S$ | $Re_S$ | $V_t$ | $\mu_S$ | $Re_S$ | $V_t$ | $\mu_S$ | $Re_S$ |
|---|---|---|---|---|---|---|---|---|---|---|---|---|
| 4.000 | -- | -- | -- | 0.046 | 1.300 | 0.137 | 0.034 | 1.760 | 0.074 | 0.078 | 0.726 | 0.542 |
| 5.000 | 0.072 | 1.289 | 0.272 | 0.065 | 1.437 | 0.219 | 0.041 | 2.241 | 0.090 | 0.113 | 0.787 | 0.905 |
| 7.000 | 0.123 | 1.504 | 0.557 | 0.103 | 1.795 | 0.391 | 0.051 | 3.643 | 0.095 | 0.185 | 0.959 | 1.702 |
| 9.000 | 0.171 | 1.759 | 0.851 | 0.134 | 2.239 | 0.525 | 0.044 | 6.868 | 0.056 | 0.255 | 1.130 | 2.560 |
| 11.000 | 0.215 | 2.088 | 1.102 | 0.152 | 2.947 | 0.553 | 0.028 | 16.230 | 0.018 | 0.315 | 1.365 | 3.199 |
| 14.280 | 0.258 | 2.928 | 1.223 | 0.156 | 4.857 | 0.446 | -- | -- | -- | 0.383 | 1.889 | 3.645 |
| 17.450 | 0.272 | 4.105 | 1.125 | 0.126 | 9.000 | 0.238 | -- | -- | -- | 0.405 | 2.639 | 3.374 |
| 20.620 | 0.250 | 6.306 | 0.795 | 0.077 | 20.400 | 0.076 | -- | -- | -- | 0.357 | 4.229 | 2.193 |

**4.2.2 On trends in terminal velocity data:** In the following we interpret the trends terminal velocity data given in section 4.2.1 in terms of Eq. 19 and Eq. 16. We use the following relation which follows from the definition of $f^V$

$$V_t = f^V(Re_\infty, \lambda) V_\infty \qquad (27)$$

(a) Fixed $D$, varying $d$: In an experiment set on $V_t$ with increasing $d$ at a fixed $D$, down a column of $V_t$ values in Table 2, both $Re_\infty$ and $\lambda$ increase. It is known from experiments (F&W, K&D) and simulations (Wham [41]) that $f^V$ decreases with increasing $\lambda$ for a fixed $Re_\infty$ (Fig. 3). With increasing $Re_\infty$ and increasing $\lambda$ variation of $f^V$ is not monotonic, $Re_\infty$ and $\lambda$ work in opposition. However, of the two, $\lambda$ dominates and in the limited range that we have covered in a group of experiments with varying $d$ at a fixed $D$, $f^V$ shows a decrease as $d$ increases in this set.

We use the result that $Re_\infty/d$ is proportional to $V_\infty$ (Eq. 3) and that $Re_\infty/d$ is an increasing function of $d$ (section 4.2.1) to infer that $V_\infty$ is an increasing function of $d$. We put the dependence of $f^V$ and $V_\infty$ on $d$ together and Eq. 27 to infer that $V_t$ will show a maximum at an intermediate value of $d$, as noted in sec.4.2.1.

(b) Fixed $d$, varying $D$: In this set, $\lambda$ changes but not $Re_\infty$. implying a constant $V_\infty$ (Eq. 3). With decreasing D and $\lambda$ and a constant $Re_\infty$ the value of $f^V$ decreases (subsection (a) above).We infer ( Eq. 27) $V_t$ decreases with increasing $\lambda$ for a fixed $d$,as noted in 4.2.1.

(c) Fixed $\lambda$, varying $d$ and $D$**:** In this set $Re_\infty$ changes but not $\lambda$. We note that $Re_\infty$ increases with increasing $d$ (section 4.1.1). At a fixed $\lambda$, an increasing $Re_\infty$ implies an increasing $f^V$. We have noted that $V_\infty$ is also an increasing function of $d$. It follows from Eq. 27 that $V_t$ increases with increasing $d$ at a fixed $\lambda$**.** System pairs in Table 2 that illustrate this feature are; $(d, D)$: $(7, 40)$ and $(5, 30)$; $(9, 40), (7, 30)$ and $(4, 16.5)$; $(11, 40), (9, 30)$ and $(5, 16.5)$.

**4.2.3 Accuracy of terminal velocity data:** The accurate measurement of coordinates and time, performed in the video camera, imparts high accuracy to the values of terminal velocity we report. A possible source of significant systematic error has however been identified in the literature. An off-axis movement of a ball falling vertically along the centerline has been considered a possibility, at very low $Re$ $(10^{-3} < Re < 10^{-1})$, on the basis of theoretical arguments and some experimental reports (Ambari et. al. [17], [47]; Christopherson and Dawson [55], Bungay and Brenner [56]). At a larger $Re$, a $Re$ dependent horizontal force (Shinohara and Hasimoto [16]) operates to restore any deviation away from an axial fall back to centerline. Off-axis movement at a very low $Re$ is a consequence of lower frictional force along a vertical axis away from the tube axis and the principle of minimum dissipation. This principle requires that the ball falls along the line of minimum vertical frictional force. Ambari et. al. [17] have shown by use of a magnetic rheometer that the eccentricity dependence of the frictional force that impedes the vertical fall shows a minimum at an intermediate eccentricity, whose location shifts to larger values, as $\lambda$ increases (from 0.08 to 0.6). Experimental evidence for this very low $Re$ off-axis fall has been reported for very large $\lambda$, when the ball nearly fills the tube (Bungay and Brenner [55]). As the radius ratio approaches unity, the force ratio should tend to infinity because of increased shear stresses between surfaces in relative motion. An additional effect arising from 'back flow' operates; as the sphere moves downwards, an equivalent amount of fluid moves upwards in the restricted annulus between the sphere and the cylinder (Ambari et.al.[45]). Presumably with decreasing $\lambda$, these effects decrease, but may not disappear. Error contamination of measured value of $V_t$ arising from an undetected off-axis movement will make it larger than what it would be for a perfectly axial fall, causing $f^V$ to be smaller than its actual value. This line of argument has been used to account for a larger $f^V$ measured in falling ball experiments of F&W and McNowan [37] than that in magnetic rheometric measurements (Ambari et. al. [17], [47]). The latter agree with 'exact theory', but the former do not [47].

In our experiments the smallest Reynolds number, $Re$ is 0.13, and the largest radius ratio is 0.667; both lie outside the range Bungay and Brenner [55] cite in their paper ($10^{-3} < Re < 10^{-1}$, $\lambda \sim 0.992$) on experimental reports on off-axis movement of a falling ball. Ambari et. al. [47] 'suspect' such an off-center fall at smaller $\lambda (< 0.6)$ but $Re$ in their report ($10^{-3}$) is still much smaller than in our experiment. Deviation from axial fall is therefore not anticipated and is also not observed in our experiments in the videos we record. In the early falling ball experiments of e.g. F&W the ball detection method was not capable of detecting off-center movements and could have gone undetected. This is not so in the video detection of falling ball trajectory that we use. We conclude that the terminal velocity values reported in Table 2 are free from any error due to possible off-center movement of the falling ball.

The videos of a falling ball in our experiments show tiny fluctuations around the vertical centerline. These off-center movements are not the ones Ambari et.al. [17], [47] refer to. We estimate error due to these off-centre fluctuations in the next section and find them to be of negligible significance.

**4.2.4 Verticality and off-center movements:** We have estimated the error in $V_t$ due to off-axis fluctuations of the falling ball trajectory as shown in Fig. 1 as follows. We calculate RMS deviation in position of the falling ball from the center line. We assume that the ball trajectory deviates off-center all the way at an eccentricity ($e_\sigma$) specified by the $\sigma$ of horizontal deviation. We equate the frictional force for a centerline fall to that of an eccentric fall with eccentricity $e_\sigma$. Both frictional forces equal $F_D$, the sum of the forces of gravity and buoyancy. The formulae for frictional forces in eccentric location of the ball are given by Tozeren [14]. For quiescent fluid and non rotating sphere $V$ and $\Omega$ of Tozeren's Eq. 5.1 are zero. The equation reduces to

$$F_{eccentric} = 6\pi\mu r(\lambda V_t) \tag{28a}$$

$$\lambda = \lambda_0 + e_\sigma \lambda_1 + e_\sigma^2 \lambda_2 + \cdots \tag{28b}$$

Our $V_t$ is Tozeren's $U$; we also drop the superscript $U$ on $\lambda$. We designate centerline $V_t$ as $V_t^C$ and off center $V_t$ as $V_t^e$. For $e_\sigma = 0$ $F_{eccentric}$ will become $F_{centerline} = 6\pi\mu r(\lambda_0 V_t^c)$. We define $\Delta V_t^{err} = V_t^{exact} - V_t^{expt}$, recognize $V_t^{exact}$ as $V_t^c$, $V_t^{expt}$ as $V_t^e$ and obtain

$$\Delta V_t^{err}/V_t^{expt} = e_\sigma^2 \lambda_2/\lambda_0 \tag{29}$$

The values of $\lambda_0$ and $\lambda_2$ are given in Tozeren [14]. The values of $e_\sigma$ for $\lambda = 0.133, 0.367$ and $0.582$ are 0.0011, 0.0014 and 0.0023 respectively and the corresponding values for $\Delta V_t^{err}/V_t^{expt}$ are -2.42E-07, -1.74E-06 and -1.84E-05 respectively. The deviations are within error of measurement of $V_t$.

**4.2.5 $\mu_S$ and $Re_S$: values and trends** : We report in Table 2 values of $\mu_S$, calculated from those of $V_t$ using Eq. 3, and those of $Re_S$, calculated from those of $\mu_S$ and $V_t$ by Eq.5. In the following we interpret trends in these values.

We infer from a restructured Eq. 13a

$$Re_S = (f^V)^2 \, (Re_\infty/(F_D/F_S)_{\lambda\to 0}) \tag{30}$$

that $Re_S$ is a function of both the ball diameter ($d$) and the tube diameter ($D$) for a given liquid. On the RHS, $(f^V)^2$ a function of $Re_\infty$ and $\lambda$, the second factor in bracket is a function of $Re_\infty$ alone and $Re_\infty$ is a function of $d$.

The dependence of $Re_S$ on $d$ and $D$ can be related to that of $V_t$ by noting that $Re_S$ is proportional to $(V_t/d)^2$.

(a) $D$ fixed, $d$ changes: $V_t$ goes through a maximum with increasing $d$, so does $V_t/d$ and $Re_S$

(b) $d$ fixed, $D$ changes: For a fixed $d$ the pattern follows that of $V_t$, a monotonic decrease at a fixed $d$ as $D$ decreases.

(c) At a nearly equal $\lambda$ (at two different $d$, $D$ pairs) we assess the variation of $Re_S$ with $d$ (and $D$) as follows: (i) $Re_\infty/(F_D/F_S)_{\lambda\to 0}$ monotonically increases with increasing $Re_\infty$ for $Re_\infty \leq 4363$, (ii) $Re_\infty$ is a monotonically increasing function of $d$, (iii) therefore, $Re_\infty/(F_D/F_S)_{\lambda\to 0}$ monotonically increases with increasing $d$ for $Re_\infty \leq$ 4363, (iv) $f^V$ is a monotonically increasing function of $Re_\infty$ and therefore of $d$ (v) as a result $Re_S$ monotonically increases with $d$ in the low $Re_\infty$ regime for equal value of $\lambda$. Values in Table 2 confirm these conclusions.

$\mu_S$ is proportional to $d^2/V_t$ for a given liquid and density of ball material. As $D$ is decreased for a given $d$, $V_t$ monotonically decreases and we observe a monotonic increase in $\mu_S$. An increase in values of $d$ for a given $D$ causes a monotonic increase in $\mu_S$ overshadowing the non-monotonic variation of $V_t$, observed with increasing $d$.

Relative errors in $\mu_S$ and $Re_S$ are of the same order as that in $V_t$.

Values of $V_t$ forms primary data set; $\mu_S$ and $Re_S$ atre quantities calculated with $V_t$ alone. In the following we discuss values of quantities, calculated with further use of Eq. 16 and Eq. 19.

**4.2.6 Reynolds number $Re_\infty$:** In Table 3 we give values of $Re_\infty^\mu$ (sec.3.4.1) of Silicone oil and Glycerol, $Re_\infty^{V_t}$ (KDC), $Re_\infty^{V_t}$ (Wham) (sec. 3.4.3) of Silicone oil with $D = 40, 30, 16.5 mm$ and all 8 ball diameters and $Re_\infty^{V_t}$ of Glycerol with $D = 40 mm$ and all 8 ball diameters.

**Table 3: Values of $Re_\infty$ (low $Re_\infty$ systems)**

| D (mm) | Silicone Oil | | | | | | | Glycerol | | |
|---|---|---|---|---|---|---|---|---|---|---|
| | $Re_\infty^\mu$ | $Re_\infty^{V_t}$ (KDC) | | | $Re_\infty^{V_t}$ (Wham) | | | $Re_\infty^\mu$ | $Re_\infty^{V_t}$ (KDC) | $Re_\infty^{V_t}$ (Wham) |
| | | D=40mm | D=30mm | D=16.5mm | D=40mm | D=30mm | D=16.5mm | D=40mm | D=40mm | D=40mm |
| 4.000 | 0.221 | -- | 0.233 | 0.259 | -- | 0.256 | 0.271 | 0.728 | 0.743 | 0.804 |
| 5.000 | 0.424 | 0.428 | 0.440 | 0.469 | 0.473 | 0.482 | 0.497 | 1.351 | 1.384 | 1.463 |
| 7.000 | 1.107 | 1.113 | 1.139 | 1.192 | 1.206 | 1.215 | 1.344 | 3.276 | 3.332 | 3.332 |
| 9.000 | 2.127 | 2.237 | 2.335 | 1.924 | 2.313 | 2.353 | 2.546 | 5.938 | 6.555 | 6.195 |
| 11.000 | 3.532 | 3.921 | 3.968 | 1.947 | 3.798 | 3.703 | 4.009 | 9.445 | 10.821 | 9.621 |
| 14.280 | 6.602 | 7.593 | 7.835 | -- | 6.412 | 6.189 | -- | 16.878 | 19.964 | 14.989 |
| 17.450 | 10.387 | 12.790 | 11.384 | -- | 8.948 | 7.903 | -- | 25.855 | 31.301 | 17.941 |
| 20.620 | 15.197 | 18.067 | 11.430 | -- | 10.314 | 8.744 | -- | 37.148 | 39.778 | 16.744 |

The values of $Re_\infty^\mu$ in Table 4 increase monotonically with increasing $d$, as an accurate estimate of $Re_\infty$ should. This feature is observerd in Fig. 5b. This trend is preserved in $Re_\infty^{V_t}$ (KDC and Wham) because the errors in $Re_\infty^{V_t}$ are not large.

$Re_\infty$ is characteristic of a ball liquid pair and is independent of tube diameter. If the equations used for calculations are exact and if $V_t^{expt}$ is accurate then a $Re_\infty^{V_t}$ triplet for a given ball liquid pair and three different tube diameters should have equal values, equal to $Re_\infty^\mu$. Data in Table 3 show that (i) $Re_\infty^{V_t}$ triplets, $D = 40, 30, 16.5 mm$, for all eight ball diameters for the same equation, KDC or Wham are unequal, i.e., tube diameter dependent, within KDC set and Wham set separately; (ii) the two sets are not equal to each other; (iii) neither set is equal to $Re_\infty^\mu$. Clearly an error is indicated. In view of the high accuracy of the values of $V_t$ (sec.4.2.3), we conclude that it is necessary to assume that the functional forms of the equations used, e.g., Eq. 16, 19, 20 may have some inexactness. An inexact form of $f^V(Re_\infty,\lambda)$ and $(F_D/F_S)_{\lambda\to 0}$ alters the functions $Re_S/(f^V)^2$ and $Re_\infty/(F_D/F_S)_{\lambda\to 0}$ and changes the point of intersection in the graphical method of determination of $Re_\infty$ (sec. 3.4.2, Fig. 4). In sec. 4.4, we introduce small modification in Eq. 19 to remove these anomalies.

**4.2.7 Bulk viscosity coefficient $\mu_\infty$ :** Values of $\mu_\infty$ of Silicone oil and Glycerol calculated from those of $V_t$ by use of KDC and Wham equations (designated $\mu_\infty^{KDC}$ and $\mu_\infty^W$ respectively) are given in Table 4. The most noticeable feature of these data sets is that $\mu_\infty$, a liquid specific ball-tube independent system property appears to vary with $d$ and $D$. An error is indicated.

**Table 4: Values of $\mu_\infty$ (low $Re_\infty$ systems)**

| d (mm) | Silicone Oil | | | | | | | | | Glycerol | | |
|---|---|---|---|---|---|---|---|---|---|---|---|---|
| | $\mu_\infty$ (KDC) | | | $\mu_\infty$ (DC) | | | $\mu_\infty$ (WC) | | | $\mu_\infty$ (KDC) | $\mu_\infty$ (DC) | $\mu_\infty$ (WC) |
| | D=40 | D=30 | D=16.5 | D=40 | D=30 | D=16.5 | D=40 | D=30 | D=16.5 | D=40 | D=40 | D=40 |
| 4.000 | -- | 0.983 | 0.932 | -- | 0.914 | 0.910 | -- | 0.935 | 0.907 | 0.598 | 0.527 | 0.571 |
| 5.000 | 1.004 | 0.989 | 0.956 | 0.910 | 0.909 | 0.939 | 0.952 | 0.942 | 0.927 | 0.594 | 0.505 | 0.576 |
| 7.000 | 1.005 | 0.992 | 0.968 | 0.877 | 0.896 | 0.951 | 0.961 | 0.957 | 0.905 | 0.596 | 0.479 | 0.596 |
| 9.000 | 0.979 | 0.955 | 1.068 | 0.838 | 0.874 | 1.030 | 0.960 | 0.951 | 0.908 | 0.565 | 0.446 | 0.586 |
| 11.000 | 0.946 | 0.939 | 1.433 | 0.821 | 0.911 | 1.200 | 0.964 | 0.979 | 0.933 | 0.550 | 0.443 | 0.595 |
| 14.280 | 0.921 | 0.904 | -- | 0.878 | 1.030 | -- | 1.027 | 1.053 | -- | 0.536 | 0.473 | 0.653 |
| 17.450 | 0.877 | 0.956 | -- | 0.965 | 1.260 | -- | 1.112 | 1.216 | -- | 0.526 | 0.536 | 0.776 |
| 20.620 | 0.896 | 1.222 | -- | 1.160 | 1.500 | -- | 1.308 | 1.459 | -- | 0.573 | 0.729 | 1.051 |

Values of $\mu_\infty^{KDC}$ of Silicone oil show a monotonic decrease with increasing $\lambda$ (as $d$ varies) in $D = 40 mm$ till it reaches a minimum beyond which an increase is observed. $\mu_\infty^{KDC}$ of Glycerol in $D = 40 mm$ tube shows essentially the same pattern. In $D = 30 mm$ with Silicone oil a small initial increase followed by a decrease to a minimum at $d = 14.28 mm$ is observed. The minimum is followed by a rise at $d = 17.45, 20.62 mm$. The minima in the two tubes appear at comparable values of $\lambda$. At the $\lambda$ values where a minimum in $\mu_\infty$ is observed $V_t$ assumes a maximum value. In $D = 16.5 mm$, the minimum in $\mu_\infty^{KDC}$ disappears, a monotonic increase in $\mu_\infty^{KDC}$ with increasing $\lambda$ is observed. The disappearance of the minimum has its counterpart in a very small maximum in $V_t$ obtained in this narrowest tube.

The pattern of dependence of $\mu_\infty^W$ of Silicone oil and Glycerol on $d$ and $D$ is different. In all three tubes with Silicone oil an overall increase which is not monotonic is observed. The same pattern is observed in $\mu_\infty^W$ of Glycerol with $D = 40mm$ when $d$ is varied. The variation of $\mu_\infty^{DC}$ with increasing $d$ at a fixed $D$ follows the same pattern as $\mu_\infty^{KDC}$ except that the minima is shifted to a smaller value of $d$.

The error indicated in the lack of constancy of $\mu_\infty$ does not arise from $V_t$; (i)if we assume that an off axis movement of the falling ball makes $V_t^{expt} > V_t^{exact}$, as is 'suspected' at $Re_\infty \to 0$ (Ambari et. al. [47]) we have $Re_\infty^{V_t} > Re_\infty^\mu$ and $\mu_\infty^{V_t} < \mu_\infty^{exact}$, inequalities that do not hold in all entries in Tables 3 and 4; (ii) we have consicered and ruled out off-axis movement of the falling ball in our experiments (sec 4.2.3, 4.2.4). We conclude that lack of constancy of $\mu_\infty$ and the inequality $Re_\infty^{V_t} \neq Re_\infty^\mu$ are entirely due to inexact functional form of KDC, DC and Wham equations.

**4.2.8 Viscosity ratio and Velocity ratio**: Tables 5 and 6 give values of these two ratios ($f^V, f^\mu$) calculated by KDC (sec. 3.4.2) and Wham (sec. 3.4.3) equation respectively as a function of their two determinants $\lambda$ and $Re_\infty$. These values for $d = 14.28mm$ and $D = 16.5mm$ with Wham equation are not reported because $\lambda$ exceeds 0.7, the upper limit of the range of validity of Wham equation.

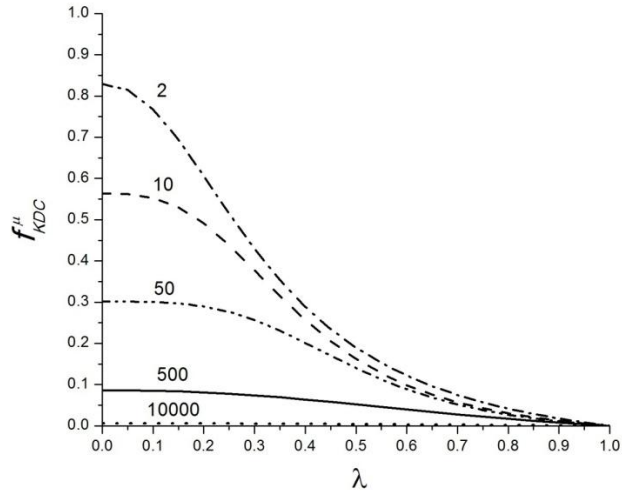

**Figure 6: Functional dependence of $f^\mu$ on $\lambda$ as given by KDC equation for a wide range of $Re_\infty$.**

**Table 5:** $f_{KDC}^v, f_{KDC}^\mu$ from values of $V_t$ alone (low $Re_\infty$ systems)

| d (mm) | Silicone Oil | | | | | | Glycerol | |
|---|---|---|---|---|---|---|---|---|
| | D=40mm | | D=30mm | | D=16.5mm | | D=40mm | D=40mm |
| | $f_{KD}^V$ | $f_{KDC}^\mu$ | $f_{KD}^V$ | $f_{KDC}^\mu$ | $f_{KD}^V$ | $f_{KDC}^\mu$ | $f_{KD}^V$ | $f_{KDC}^\mu$ |
| 4.000 | -- | -- | 0.778 | 0.758 | 0.543 | 0.528 | 0.889 | 0.822 |
| 5.000 | 0.817 | 0.780 | 0.724 | 0.690 | 0.450 | 0.428 | 0.875 | 0.775 |
| 7.000 | 0.749 | 0.669 | 0.622 | 0.554 | 0.300 | 0.266 | 0.827 | 0.638 |
| 9.000 | 0.684 | 0.557 | 0.528 | 0.427 | 0.187 | 0.156 | 0.809 | 0.541 |
| 11.000 | 0.621 | 0.453 | 0.438 | 0.319 | 0.106 | 0.089 | 0.768 | 0.448 |
| 14.280 | 0.512 | 0.315 | 0.306 | 0.186 | -- | -- | 0.678 | 0.324 |
| 17.450 | 0.412 | 0.214 | 0.197 | 0.106 | -- | -- | 0.572 | 0.234 |
| 20.620 | 0.310 | 0.142 | 0.111 | 0.060 | -- | -- | 0.415 | 0.157 |

Dependence of $f^V$ and $f^\mu$ on $\lambda$ and $Re_\infty$ as given by KDC equation, are shown in Fig. 3 and 6 respectively. In Fig. 3 we observe an increase in $f^V$ at a fixed $\lambda$ with an increase in $Re_\infty$ and a decrease in $f^V$ at a fixed $Re_\infty$ with increase in $\lambda$. The plots become increasingly insensitive to changes in $Re_\infty$ as $Re_\infty$ becomes large (100 upwads). The $\lambda$ dependence of $f^\mu$ is derived entirely from that of $f^V$ since $(F_D/F_S)_{\lambda \to 0}$ is $\lambda$ independent (Eq. 16 and 17). $f^\mu(\lambda)$ at different fixed values of $Re_\infty$ is shown in Fig. 6.

**Table 6:** $f_w^v, f_w^\mu$ from values of $V_t$ alone (low $Re_\infty$ systems)

| d (mm) | Silicone Oil | | | | | | Glycerol | |
|---|---|---|---|---|---|---|---|---|
| | D=40mm | | D=30mm | | D=16.5mm | | D=40mm | D=40mm |
| | $f_{WC}^V$ | $f_W^\mu$ | $f_{WC}^V$ | $f_W^\mu$ | $f_{WC}^V$ | $f_W^\mu$ | $f_{WC}^V$ | $f_W^\mu$ |
| 4.000 | -- | -- | 0.744 | 0.723 | 0.532 | 0.516 | 0.858 | 0.788 |
| 5.000 | 0.779 | 0.739 | 0.693 | 0.657 | 0.438 | 0.415 | 0.846 | 0.732 |
| 7.000 | 0.723 | 0.640 | 0.604 | 0.534 | 0.285 | 0.249 | 0.822 | 0.622 |
| 9.000 | 0.674 | 0.546 | 0.526 | 0.425 | 0.166 | 0.132 | 0.797 | 0.519 |
| 11.000 | 0.628 | 0.462 | 0.450 | 0.333 | 0.079 | 0.058 | 0.763 | 0.436 |
| 14.280 | 0.544 | 0.351 | 0.333 | 0.217 | -- | -- | 0.704 | 0.346 |
| 17.450 | 0.464 | 0.271 | 0.223 | 0.135 | -- | -- | 0.640 | 0.294 |
| 20.620 | 0.372 | 0.208 | 0.122 | 0.072 | -- | -- | 0.527 | 0.249 |

The values of $f^\mu(Re_\infty, \lambda)$ become progressively smaller at all $\lambda$ as $Re_\infty$ increases. The increase in $f^V(Re_\infty, \lambda)$ at all $\lambda$ with increase in $Re_\infty$ is more than offset by a $\lambda$ independent $(F_D/F_S)_{\lambda \to 0}$ (Eq. 16 and 17) that is greater than unity and monotonically increases with $Re_\infty$. At larger $Re_\infty$, values of $f^\mu$ are less sensitive to $\lambda$ as is apparent in the progressively flatter appearance of the plots in Fig. 6. Increase in $\lambda$ and increase in $Re_\infty$ have opposing effects on values of $f^V$ but not on values of $f^\mu$. Over the range of $\lambda$ and $Re_\infty$ values that we have investigated a decrease in $Re_\infty$ for $f^\mu$ in no case annuals the effect of increasing $\lambda$.

The trends of values in Tables 5 and 6 are consistent with those of Fig. 3 and 6 referred to above. In Tables 5 and 6, down each column (fixed $D$) $\lambda$ and $Re_\infty$ both increase. The effect of $\lambda$ dominates and $f^V$ shows a monotonic decrease. At concordant values of $\lambda$ and $Re_\infty$ in tubes of different diameters values of $f^V$ and $f^\mu$ are closely spaced. At the largest $\lambda$ and $Re_\infty$ the values of $f^V$ and $f^\mu$ are quite small (e.g. = 0.679, $Re_\infty^\mu$ = 3.532, $f^V$ = 0.083 and $f^\mu$ = 0.062), implying significant corrections to $\mu_S$. In our discussion on correction factors at much larger $Re_\infty$ we find yet smaller $f^\mu$ (viz., 0.003). Increase of $Re_\infty$ by three orders of magnitude at comparable $\lambda$ increases $f^V$, and decreased $f^\mu$.

In Table 7 we give values of $f^V(\mu, V_t)$ and $f^\mu(\mu, V_t)$ calculated by procedure given in sec. 3.4.6. These values are more accurate than those in Table 5 and 6 since the former have not inherited the effect of inexact functional form of Eq. 19 and 20. The trends in dependence of $f^V(\mu, V_t)$ and $f^\mu(\mu, V_t)$ on $\lambda$ and $Re_\infty$ in this table are the same as that discussed for $f_{KD}^V(V_t)$ and $f_W^V(V_t)$ in Tables 5 and 6. This is so because the differences $(f^V(\mu, V_t) - f_{KD}^V(V_t))$ are small.

**Table 7: $f^V$, $f^\mu$ calculated from values of $V_t$, $\mu_\infty$ (low $Re_\infty$ systems)**

| d (mm) | Silicone Oil | | | | | | Glycerol | |
| --- | --- | --- | --- | --- | --- | --- | --- | --- |
| | D=40mm | | D=30mm | | D=16.5mm | | D=40mm | |
| | $f^V(\mu,V_t)$ | $f^\mu(\mu,V_t)$ | $f^V(\mu,V_t)$ | $f^\mu(\mu,V_t)$ | $f^V(\mu,V_t)$ | $f^\mu(\mu,V_t)$ | $f^V(\mu,V_t)$ | $f^\mu(\mu,V_t)$ |
| 4.000 | -- | -- | 0.797 | 0.778 | 0.586 | 0.571 | 0.896 | 0.829 |
| 5.000 | 0.819 | 0.782 | 0.735 | 0.702 | 0.471 | 0.450 | 0.875 | 0.765 |
| 7.000 | 0.751 | 0.670 | 0.629 | 0.562 | 0.310 | 0.277 | 0.827 | 0.628 |
| 9.000 | 0.698 | 0.573 | 0.548 | 0.450 | 0.179 | 0.147 | 0.809 | 0.533 |
| 11.000 | 0.646 | 0.483 | 0.458 | 0.342 | 0.083 | 0.062 | 0.768 | 0.441 |
| 14.280 | 0.538 | 0.344 | 0.325 | 0.208 | -- | --- | 0.678 | 0.319 |
| 17.450 | 0.441 | 0.246 | 0.202 | 0.112 | -- | -- | 0.572 | 0.228 |
| 20.620 | 0.327 | 0.160 | 0.101 | 0.049 | -- | -- | 0.415 | 0.142 |

**4.2.9 $\Delta f^V$ and $\Delta f^\mu$:** The numerical values of error in calculated $f^V$ namely $\Delta f_{KD}^V$ and $\Delta f_W^V$ are defined in Eq. 23a and are calculated using data in Table 5 to 7. $\Delta f^V$ can be split into two components $\Delta f_1^V$ and $\Delta f_2^V$ as defined in Eqs. 23c and 23d respectively. The error in $f^\mu$ ($\Delta f^\mu$) is calculated by the method specified in sec. 3.4.7. In Table 8 we give $\Delta f^V$, $\Delta f^\mu$, $\Delta f^V/f^V$ and $\Delta f^\mu/f^\mu$ for KDC. In the calculation of these ratios and of the error terms $\Delta f^V$ and $\Delta f^\mu$, we use $f^V(\mu, V_t)$ and $f^\mu(\mu, V_t)$ (Table 7) as reference. The values of $f^\mu(\mu, V_t)$ are as accurate as experimental values of $\mu_\infty$, $V_t$; $f^V(\mu, V_t)$ inherit any deviation that Eq. 16, may have from 'true' values.

$\Delta f_{KDC}^V$ in a $40mm$ tube is small, first significant figure being in third decimal place, for small $d$, rises to a maximum, to decrease again at a larger $d$. In $30mm$ tube, we observe a decrease followed by an increase and then a decrease once again. In the larger $d$ end, the sign becomes negative. In $16.5mm$ tube, a monotonic decrease is observed, with a switch to negative sign beyond which its magnitude keeps increasing. $\Delta f_W^V$ values show a decrease with increase in $d$ ($D$ fixed), a changeover to a negative sign. The negative values show an increase in magnitude with further increase in $d$. In 16.5mm tube, the negative values are not observed. Mostly, K-D equation shows a larger deviation from 'correct' value of $f^V$ at intermediate $\lambda$ values. In contrast, Wham equation shows a smaller deviation at intermediate $\lambda$ values, the deviation increases with different signs on two sides of this 'intermediate' range. These complex features arise from effects of varying $\lambda$ and $Re_\infty$.

$\Delta f^\mu$ in all but two entries(in both KDC and Wham) in Table 8 (the last two entries in Glycerol data) are larger in magnitude than $\Delta f^V$. $f^\mu < f^V$ in all cases, since $(F_D/F_S)_{\lambda \to 0} > 1$ (Eq. 11). As a result, $\frac{\Delta f^\mu}{f^\mu} > \frac{\Delta f^V}{f^V}$ in all cases, even in the ones in which $\Delta f^\mu$ is $< \Delta f^V$. In KDC calculations with Silicone oil the values of $\frac{\Delta f^\mu}{f^\mu}$ exceed 5% in 3/5 (D=16.5), 4/8 (D=30). 4/7 (D=40) entries and in those with Glycerol (D=40), in 4/8 entries, in Table 8. They exceed 10% in 1/5 ($D = 16.5$), 2/8 ($D = 30$), 2/7 ($D = 40$) and in 2/8 (Glycerol, $D = 40$) entries. All of these 7/28 entries with $(\Delta f^\mu/f^\mu) > 10\%$, have larger $\lambda$ (~0.5). In Wham calculations the errors are larger in 19/28 entries, more significantly so at large $\lambda$ and are smaller in the rest, which lie in the intermediate $\lambda$ range, viz. $\lambda = 0.225 - 0.357$ in $D = 40mm$, Glycerol; $\lambda = 0.275 - 0.436$ in $D = 40mm$, Slicone oil, where a switch over from positive to negative error values is observed. Clearly large $\lambda$ with accompanying large $Re_\infty$ values increase error in $\Delta f^\mu/f^\mu$ and increase relative error in $\mu_\infty$. The accuracy of Wham equation at intermediate $\lambda$ values is noted and is referred to later.

Translation of $\Delta f^V \to \Delta f^\mu$ proceeds as given in sec.3.4.7.2. In Table 9 we observe that both $\Delta f_1^V/f^V$ and $\Delta f_2^V/f^V$ increase as one increases $d$ at a fixed $D$ and that $\Delta f_2^V/f^V$ is significantly larger. At a small $Re_\infty$ $\Delta f_1^V/f^V$ and $\Delta f_1^\mu/f^\mu$ can have either relative sign and are both small in value (sec.3.4.7.2).This is observed in Table 9. The major contributor to $\Delta f^\mu/f^\mu$ is $\Delta f_2^\mu/f^\mu$. $\Delta f_2^\mu/f^\mu$ equals $\Delta f_2^V/f^V$ in all cases, as it should be, by Eq. 24f.

The magnitude and sign of $\Delta f_2^V$ result from inexactness of functional form of correlation formula being used and is expected to show no systematic variation with $\lambda$, $Re_\infty$. Smaller value of $\Delta f_1^V$,(compared to $\Delta f_2^V$) results from insensitivity of $f^V$ to change in $Re_\infty$ (sec. 3.3.4, Fig.3). The sign of $\Delta f_2^V$ is opposite of that of $\Delta f_1^V$ in all cases. $\Delta f_1^V$ is negative and small in all cases except three, in which it is positive and small. In these three cases $Re_\infty^{V_t} < Re_\infty^\mu$. As a result mostly, $|\Delta f_2^V| < |\Delta f^V|$, remaining comparable and the pattern of variation of $\Delta f_2^V$ with $d$ (for a fixed $D$) is the same as that of $\Delta f^V$. Lack of systematic variation of $\Delta f_2^V$ with $\lambda$, $Re_\infty$ makes variation of $\Delta f^V$ complex in Table 8.

Table 8: Error estimates in $f^V_{KDC}$ and $f^\mu_{KDC}$ in low $Re_\infty$ systems

| | | KDC | | | | Wham | | | |
|---|---|---|---|---|---|---|---|---|---|
| | $\lambda$ | $\Delta f^\mu$ | $\Delta f^\mu/f^\mu$ | $\Delta f^V$ | $\Delta f^V/f^V$ | $\Delta f^\mu$ | $\Delta f^\mu/f^\mu$ | $\Delta f^V$ | $\Delta f^V/f^V$ |
| Silicone Oil D=16.5mm | 0.242 | 0.044 | 0.076 | 0.042 | 0.072 | 0.055 | 0.097 | 0.054 | 0.092 |
| | 0.303 | 0.022 | 0.050 | 0.021 | 0.045 | 0.035 | 0.078 | 0.033 | 0.071 |
| | 0.424 | 0.011 | 0.038 | 0.010 | 0.031 | 0.028 | 0.100 | 0.025 | 0.081 |
| | 0.545 | -0.009 | -0.062 | -0.008 | -0.046 | 0.014 | 0.098 | 0.013 | 0.070 |
| | 0.667 | -0.027 | -0.427 | -0.023 | -0.281 | 0.004 | 0.071 | 0.004 | 0.045 |
| | $\lambda$ | $\Delta f^\mu$ | $\Delta f^\mu/f^\mu$ | $\Delta f^V$ | $\Delta f^V/f^V$ | $\Delta f^\mu$ | $\Delta f^\mu/f^\mu$ | $\Delta f^V$ | $\Delta f^V/f^V$ |
| Silicone Oil D=30mm | 0.133 | 0.019 | 0.025 | 0.019 | 0.024 | 0.055 | 0.071 | 0.053 | 0.067 |
| | 0.167 | 0.012 | 0.017 | 0.012 | 0.016 | 0.045 | 0.064 | 0.043 | 0.058 |
| | 0.233 | 0.008 | 0.015 | 0.007 | 0.012 | 0.028 | 0.049 | 0.025 | 0.040 |
| | 0.300 | 0.023 | 0.052 | 0.021 | 0.038 | 0.025 | 0.056 | 0.022 | 0.041 |
| | 0.367 | 0.023 | 0.068 | 0.020 | 0.044 | 0.010 | 0.028 | 0.008 | 0.018 |
| | 0.476 | 0.021 | 0.102 | 0.019 | 0.057 | -0.009 | -0.045 | -0.008 | -0.025 |
| | 0.582 | 0.006 | 0.051 | 0.005 | 0.025 | -0.023 | -0.208 | -0.021 | -0.104 |
| | 0.687 | -0.011 | -0.213 | -0.010 | -0.099 | -0.022 | -0.449 | -0.021 | -0.202 |
| | $\lambda$ | $\Delta f^\mu$ | $\Delta f^\mu/f^\mu$ | $\Delta f^V$ | $\Delta f^V/f^V$ | $\Delta f^\mu$ | $\Delta f^\mu/f^\mu$ | $\Delta f^V$ | $\Delta f^V/f^V$ |
| Silicone Oil D=40mm | 0.100 | -- | -- | -- | -- | -- | -- | -- | -- |
| | 0.125 | 0.002 | 0.003 | 0.002 | 0.003 | 0.043 | 0.055 | 0.041 | 0.050 |
| | 0.175 | 0.002 | 0.002 | 0.001 | 0.002 | 0.031 | 0.045 | 0.028 | 0.037 |
| | 0.225 | 0.016 | 0.028 | 0.014 | 0.020 | 0.027 | 0.047 | 0.024 | 0.034 |
| | 0.275 | 0.030 | 0.061 | 0.026 | 0.040 | 0.021 | 0.043 | 0.018 | 0.028 |
| | 0.357 | 0.030 | 0.086 | 0.026 | 0.048 | -0.007 | -0.019 | -0.006 | -0.011 |
| | 0.436 | 0.032 | 0.130 | 0.029 | 0.066 | -0.026 | -0.104 | -0.023 | -0.052 |
| | 0.516 | 0.018 | 0.111 | 0.017 | 0.053 | -0.048 | -0.298 | -0.045 | -0.136 |
| | $\lambda$ | $\Delta f^\mu$ | $\Delta f^\mu/f^\mu$ | $\Delta f^V$ | $\Delta f^V/f^V$ | $\Delta f^\mu$ | $\Delta f^\mu/f^\mu$ | $\Delta f^V$ | $\Delta f^V/f^V$ |
| Glycerol D=40mm | 0.100 | 0.008 | 0.009 | 0.007 | 0.008 | 0.042 | 0.050 | 0.039 | 0.043 |
| | 0.125 | 0.010 | 0.013 | 0.009 | 0.010 | 0.033 | 0.043 | 0.029 | 0.034 |
| | 0.175 | 0.006 | 0.010 | 0.005 | 0.006 | 0.006 | 0.010 | 0.005 | 0.006 |
| | 0.225 | 0.032 | 0.061 | 0.028 | 0.035 | 0.014 | 0.026 | 0.012 | 0.015 |
| | 0.275 | 0.038 | 0.086 | 0.034 | 0.045 | 0.005 | 0.012 | 0.005 | 0.006 |
| | 0.357 | 0.035 | 0.109 | 0.034 | 0.051 | -0.027 | -0.085 | -0.026 | -0.039 |
| | 0.436 | 0.029 | 0.126 | 0.031 | 0.055 | -0.066 | -0.289 | -0.068 | -0.118 |
| | 0.516 | 0.007 | 0.047 | 0.008 | 0.019 | -0.106 | -0.746 | -0.113 | -0.271 |

$\Delta f_2^\mu$ is smaller in magnitude, with the same sign, than $\Delta f_2^V$ (Eq. 24). Also, $|\Delta f_2^\mu| > |\Delta f_1^\mu|$, a result of a similar inequality between the two components of $\Delta f^V$. Unlike the ($\Delta f_2^\mu$, $\Delta f_2^V$) pair, the relation between $\Delta f_1^\mu$ and $\Delta f_1^V$ is not monotonic, $\Delta f_1^\mu$ and $\Delta f_1^V$ do not always have same sign, but $\Delta f_1^\mu$ is comparatively reduced in magnitude and contributes to a minor extent to $\Delta f^\mu$. In no case $\Delta f_1^\mu$ has been amplified with respect to $\Delta f_1^V$ in the lower $Re_\infty$ regime under consideration and its sign has changed only in $40mm$ tube; its smaller magnitude makes the change of sign unimportant. The opposite sign of $\Delta f_2^V$ and $\Delta f_1^V$ in all cases reduces $\Delta f^V$ to a level below $\Delta f^\mu$, which is primarily $\Delta f_2^\mu (\Delta f_2^\mu < \Delta f_2^V)$.

Qualitatively one can understand the relative magnitude of $\Delta f^V$ and $\Delta f^\mu$ in the following terms: $\Delta f^V$ arises from an incorrect value of $Re_\infty$, namely $Re_\infty^{V_t}$ being substituted an incorrect functional form $f^V(Re_\infty, \lambda)$. $\Delta f^\mu$ arises from a substitution of $Re_\infty^{V_t}$ in an incorrect form $f^V(Re_\infty, \lambda)$ and in a more accurate $(F_D/F_S)_{\lambda \to 0}$ (Eq. 16). In this case, use of $Re_\infty^{V_t}$ introduces deviation in two places, which may act coherently or may mutually compensate. The pattern of variation of $\Delta f^V$ is maintained in that of $\Delta f^\mu$, as $(F_D/F_S)_{\lambda \to 0}$, which appears in the translation of $\Delta f^V$ to $\Delta f^\mu$ (Eq. 29) varies over a limited range in the cases we study.

### 4.3 Correlation equations, graphs and tables : Accuracy issues:

The accuracy of the calculated values of $Re_\infty$, $f^V$, $f^\mu$ and $\mu_\infty$ depend on the accuracy of functional forms of the equations used viz., KD, Wham and Cheng equations.

**4.3.1 $(F_D/F_S)_{\lambda \to 0}$: Cheng equation, Clift equation:** Cheng [40] reports the average deviation of values of $C_D$ predicted by his equation (Eq. 16) from experimental values as ~ 1.7% and a RMS deviation of 4.5%. The scatter of carefully chosen experimental values around the prediction of its predecessor Clift equation (Eq. 17), which is very close to it over the whole range is: 81.5% (±5%), 98.1% (±10%), 99.6% (±15%) (Brown and Lawler [49]). The difference between average error and RMS deviation of Cheng equation from experiments suggests existence of larger deviations in a small subset of the experimental data, as is also shown by the statistics of scatter data around Clift equation. These formulae are based on a large number of carefully chosen experimental data published by many different laboratories. The experiments being 'accurate', the errors introduced by the choice of a specific functional form are systematic. They are mostly 'small' but can be significant in the small subset responsible for a RMSD much larger than average error, e.g., 1.5% of values that deviate from Clift equation by >10% but <15% or the 0.4% that lie beyond 15%. These errors are inherent in the use of best functional form. No 'error' in the functional form is referred to. We take the form of Eq. 16 as error free.

arXiv:1202.1400 v2 [physics.flu-dyn] 13Feb 2012arXiv: 1202.1400 v2[physics.flu-dyn] 13Feb 2012**Table 8: Translation of** $\frac{\Delta f_1^V}{f^V} \to \frac{\Delta f_1^\mu}{f^\mu}$ **and** $\frac{\Delta f_2^V}{f^V} \to \frac{\Delta f_2^\mu}{f^\mu}$ **in KDC calculations in low** $Re_\infty$ **cases.**

| | $\lambda$ | $\frac{\Delta f_1^V}{f^V}$ | $\frac{\Delta (F_D/F_S)_{\lambda\to 0}}{(F_D/F_S)_{\lambda\to 0}}$ | $\frac{\Delta f_1^\mu}{f^\mu}$ | $\frac{\Delta f_2^V}{f^V} = \frac{\Delta f_2^\mu}{f^\mu}$ |
|---|---|---|---|---|---|
| Silicone Oil D=16.5mm | 0.242 | -0.007 | 0.000 | -0.003 | 0.080 |
| | 0.303 | -0.006 | -0.003 | -0.002 | 0.051 |
| | 0.424 | -0.006 | -0.007 | 0.002 | 0.037 |
| | 0.545 | 0.008 | 0.018 | -0.008 | -0.054 |
| | 0.667 | 0.043 | 0.103 | -0.103 | -0.324 |
| Silicone Oil D=30mm | 0.133 | -0.002 | 0.003 | -0.001 | 0.026 |
| | 0.167 | -0.002 | 0.000 | 0.000 | 0.018 |
| | 0.233 | -0.003 | -0.003 | 0.000 | 0.014 |
| | 0.300 | -0.011 | -0.014 | 0.004 | 0.048 |
| | 0.367 | -0.017 | -0.025 | 0.007 | 0.061 |
| | 0.476 | -0.029 | -0.053 | 0.016 | 0.087 |
| | 0.582 | -0.015 | -0.029 | 0.010 | 0.040 |
| | 0.687 | 0.054 | 0.097 | -0.061 | -0.153 |
| Silicone Oil D=40mm | 0.125 | 0.000 | 0.002 | 0.000 | 0.003 |
| | 0.175 | 0.000 | 0.000 | 0.000 | 0.002 |
| | 0.225 | -0.005 | -0.007 | 0.003 | 0.026 |
| | 0.275 | -0.014 | -0.022 | 0.008 | 0.054 |
| | 0.357 | -0.026 | -0.043 | 0.012 | 0.074 |
| | 0.436 | -0.047 | -0.070 | 0.017 | 0.113 |
| | 0.516 | -0.045 | -0.065 | 0.014 | 0.098 |
| Glycerol D=40mm | 0.100 | -0.001 | -0.002 | 0.001 | 0.009 |
| | 0.125 | -0.001 | -0.006 | 0.002 | 0.011 |
| | 0.175 | -0.002 | -0.001 | 0.002 | 0.008 |
| | 0.225 | -0.013 | -0.026 | 0.013 | 0.048 |
| | 0.275 | -0.023 | -0.045 | 0.019 | 0.067 |
| | 0.357 | -0.040 | -0.065 | 0.018 | 0.091 |
| | 0.436 | -0.059 | -0.080 | 0.012 | 0.114 |
| | 0.516 | -0.023 | -0.030 | 0.005 | 0.042 |

arXiv: 1202.1400 v2[physics.flu-dyn] 13Feb 201234

**4.3.2 $f^V(Re_\infty, \lambda)$:** Whereas formulae for $(F_D/F_S)_{\lambda \to 0}$ (Eq.16,17) are functions of $Re_\infty$ alone, those of $f^V$ is a function of two independent variables, $\lambda$ and $Re_\infty$. Number of data points, necessary for parameterizing a two-parameter function with comparable accuracy level, is considerably larger than that for a single parameter function. Regarding number of data points available for parameterization, the reverse is the case with $(F_D/F_S)_{\lambda \to 0}$ and $f^V$.

In what follows, we assess the accuracy of K-D equation (Eq. 19) in predicting experimental results. We assess the accuracy of the experimental results on velocity ratio used for deriving this formula e.g., those of F&W and those of K&D.

**4.3.2.1 Fidleris and Whitmore [36]:** These authors published an extensive experimental report on $f^V$ as a function of $Re$ (0.05 − 20,000) and $\lambda$ (0.05 – 0.6). Their $V_t$ measurements have high accuracy ( 0.3%). Uncertainty in values of $f^V$ is however significantly larger. The source of this additional uncertainty is that $V_\infty$ measurements could not have been performed with their ball position sensing system, on the same ball-liquid combination on which $V_t$ measurements were performed. The induction coil they use to sense the position of a magnetic ball falling along center-line of a cylindrical tube would not register a signal if the tube diameter is very large ($\lambda \to 0$ limit). The authors therefore use values of $C_D$ vs $Re_\infty$ reported in the literature by other investigators who measured terminal ball velocity in an infinite medium. These measurements were performed at $Re$ values, for which values of $V_t$ measured in a narrow tube were not available. In order to determine $f^V$ at a given $Re$ and $\lambda$, the authors use a family of continuous $\log C_D$ vs $\log (Re)$ plots (their $\psi$ is $C_D$) for different values of $\lambda$ (including $\lambda = 0$) and a graphical construction to evaluate $f^V$. Experimental points are scattered around these smooth lines. Two sets of correction factors, to calculate $V_t$ from $V_\infty$ ($V_t/V_\infty$) and vice versa ($V_\infty/V_t$), are reported. Their difference is an estimate of error due to scatter around the smooth curves. The difference in the value pair is ~2% for $Re = 1000, 3000, 10000$ at $\lambda = 0.6$; is larger (~10%) for $Re = 100$ as well as $Re = 0.1$ at $\lambda = 0.6$; (their Fig. 3 and 4). This uncertainty decreases with decreasing $\lambda$ for each $Re$.

Relative scarcity of experimental values of $V_t$ (narrow tube) at low and intermediate value of Reynolds number in the work of F&W prompted experimental studies in these $Re$ regimes (Sutterby [38], $Re < 4$, Kehlenbeck and DiFelice [34], $Re_\infty = 10, 20, 90$).

**4.3.2.2 Kehlenbeck and DiFelice [34]:** The authors report $f^V$ in the intermediate range of $Re_\infty$ (10 − 90) by determination of $V_t$ and $V_\infty$ for the same ball-liquid pair. Their method of terminal velocity determination use a stop watch and can be used with equal facility in a narrow tube and in a large diameter tube. The uncertainty in their $f^V$ values arises from the lower precision of their velocity measurements (~5%) and not from the use of a correlation graphs with its own inherent shortcoming, as is the case with values of F&W. Whereas F&W's errors are systematic, K-D's errors are random.

The two parameter correlation formula of an assumed functional form has been optimized by K&D with respect to their own data (all data points in their Fig. 2, $Re_\infty = 20$ and 25 out of 35 data points in Fig. 5, $Re_\infty = 10$ are their own) and the data of F&W (5 out of 10 data points at $Re_\infty = 90$ are due to F&W, rest is from K&D). The data of F&W that they use are obtained from the plots of F&W and are not primary experimental data. K&D do not estimate the accuracy with which their equation predicts the mean values of $V_t/V_\infty$. We have estimated the deviation of experimental $f^V$ data (mean values) that K&D used to deduce their equation, from the values calculated from K-D equation. RRMSD of ~30 data points in Fig. 5 ($Re_\infty = 10$) of K-D is 6.2%, but it reduces to 3% if values of $\lambda$ used in the estimate is restricted to $< 0.3$. In Fig. 6 ($Re_\infty = 90$) of K-D, RRMSD of about 10 data points is ~2.8%. Both figures have one isolated instance of large relative deviation (viz. 18%, 28%) which are so large because they belong to data points with large $\lambda$ i.e., small $f^V$. An examination of their Fig. 8 which summarizes this deviation without grouping them according to $Re_\infty$ we find some large relative deviations in points with small $f^V$. It is a fair conclusion that, with a small $\lambda$ ($< 0.3$), the relative deviation is ~$3 - 4$%. The number of data points used in this parameterization is considerably less than that used for deriving Clift or Cheng equation. K-D equation and its predecessor DiFelice equation assume a specific functional form chosen on the basis of inspection of experimental data using a fluid mechanical analogy. They find the best values of parameters of these equation forms that fit the available experimental data or by using a fluid-mechanical analogy. The resulting formula is the 'best' amongst a family and may not be best globally.

### 4.3.3 $f^\mu(Re_\infty, \lambda)$:

**4.3.3.1 Wham et.al[41]:** The authors estimate the predictive accuracy of the equation they derive (Eq. 20), i.e., the extent of agreement between computational results on drag coefficients and the values calculated from their correlation formula to be ~ 5%.. Satisfactory agreement with experiments is reported only at a specific value of $\lambda = 0.3125$ over a restricted range of $Re$ ($0 - 80$).

In view of the uncertainties discussed above, a small deviation in the correlation formula from its 'exact' counterpart is only to be anticipated. An inherent systematic error (as distinct from random measurement error) is introduced in fitting experimental data within a correlation formula. The uncertainty in fitting data is often larger than random experimental error of measurement.

**4.3.3.2 Sutterby [38]:** The author reports 160 experimental values of $\frac{\mu_s}{\mu_\infty}$ ($= \frac{1}{f^\mu}$) and interpolates between these points to work out a correlation table, for use with an unknown test liquid. Although a $\lambda$ range $0.0025 - 0.125$ was studied, fewer data points were determined at larger $\lambda$ (4 points above $\lambda = 0.10$, ~20 points above $\lambda = 0.05$) and also at larger $Re$ (88 points $Re < 0.05$). Thus although very low $\lambda$ and very low $Re$ regime are adequately covered by actual experimental points, one must rely on interpolated values as soon as $\lambda$ and $Re$ increases. The accuracy with which Sutterby's tables predict his own 'experimental values' is high, mean absolute deviation being 0.3%, but this agreement does not guarantee such a high accuracy level in those parts of ($\lambda$, $Re$) plot where actual experimental points are scarce (Sutterby's Fig. 1, 2).The process of interpolation is not specified . The $Re = 0$ limit for each $\lambda$ is tied to a specific point by use of Faxen equation (Happel and Brenner [2$^b$]. This limiting value determines the curvature of the plot at low $Re$ at each $\lambda$. The line corresponding to $\lambda = 0.125$ (the largest $\lambda$) has only 4 points, one each at $Re = 3.5$ and $2.3$ and two very closely spaced points at $Re \sim 0.2 - 0.3$. A large part of this line is experimentally unscanned and small errors may enter if a test liquid is examined in the unscanned region.

**4.3.4 Limiting forms:** The values of $K = 1/f^V$ given by K-D and Wham equations in the limit $Re$, $Re_\infty \to 0$ for different values of $\lambda$ are compared in Table 10 with theoretical estimates of $(F_D/F_S)^{-1}_{Re_\infty \to 0}$ ('exact theory' and H-S equation (Eq. 22)), with $(F_D/F_S)^{-1}_{Re_\infty \to 0}$ given in Eq. 21 (Wham equation). The limiting values determined by theoretical methods fall in two classes: (a) 'exact theory', as given in Haberman's thesis [42] that match nearly exactly with results of several accurate calculations published later by Tozeren [14], Coutanceau [44] and Paine and Sherr [43], summarized in Ambari et. al. [47] and (b) 'approximate theory', namely H-S equation and the $Re_\infty \to 0$ limiting form of Wham equation (Eq. 21). Happel and Brenner (1983$^a$) and Ambari et.al.[47] compare values of $K(=1/f^V)$ obtained from 'exact' theory and 'approximate' theory and the difference increases with increasing $\lambda$. The difference between different 'exact' theories can be ignored in our context. The exact theory is in agreement with the very careful experiments of Ambari et. al [47].

**Table 9: Different estimates of $f^V|_{Re_\infty \to 0}$**

|  | Values of $f^V$ | | | | | Relative deviation from Exact-Threory | | | |
| --- | --- | --- | --- | --- | --- | --- | --- | --- | --- |
| $\lambda$ | E-T$^*$ | H-S | K-D | W | D | H-S | K-D | W | D |
| 0.1 | 0.792 | 0.792 | 0.787 | 0.790 | 0.789 | 0.000 | 0.006 | 0.002 | 0.003 |
| 0.2 | 0.595 | 0.595 | 0.648 | 0.592 | 0.600 | 0.000 | -0.088 | 0.005 | -0.008 |
| 0.3 | 0.422 | 0.422 | 0.452 | 0.417 | 0.435 | 0.000 | -0.072 | 0.011 | -0.031 |
| 0.4 | 0.278 | 0.279 | 0.329 | 0.273 | 0.296 | -0.004 | -0.184 | 0.019 | -0.063 |
| 0.5 | 0.168 | 0.170 | 0.219 | 0.162 | 0.184 | -0.017 | -0.306 | 0.031 | -0.099 |
| 0.6 | 0.090 | 0.094 | 0.142 | 0.084 | 0.101 | -0.051 | -0.582 | 0.061 | -0.121 |

*E-T: Exact-Theory; W: Wham; D: DiFelice

The $Re_\infty \to 0$ limit of Wham equation (Eq. 21) is, by design, almost identical in form to H-S equation(Eq. 22) and shows only slightly larger deviation from 'exact' theory(Table 10). DiFelice equation (Eq. 18) has a differently structured functional form; in the $Re_\infty \to 0$ limit, however it gives values very close to those of Wham equation (Table 10).

The values of $(f_{KD}^V)_{Re \to 0}$ show larger deviation from theoretical estimates in the range $\lambda$ (0.1 − 0.6) than do the rest and are larger in value. In the design of functional forms of K-D equation and its predecessor DiFelice equation, validity over an extended $Re_\infty$ range and not in the $Re_\infty \to 0$ limit is the prime focus. The inexact $Re_\infty \to 0$ limit of K-D equation, but a more exact limit of DiFelice equation results from a modification of equation form to accommodate 'intermediate' $Re_\infty$ range in K-D equation

### 4.3.5 Comparison of different estimates of $f^\mu$ over an extended $Re$ range:

**4.3.5.1 KDC and Sutterby:** In Fig. 7a we compare values of $f^\mu$ as calculated by KDC equation and the values given in Sutterby's tables (sec. 4.3.3.2). The difference at low $Re$ ($< 1$) is consistent with the deviation of $f^V_{KD}$ from $f^V$ given by exact theory, an indication of error in K-D equation in the limit of very small $Re$ (sec 4..3.4). The agreement at a higher $Re$ ($1 < Re < 4$) indicates correctness of K-D equation in this range.

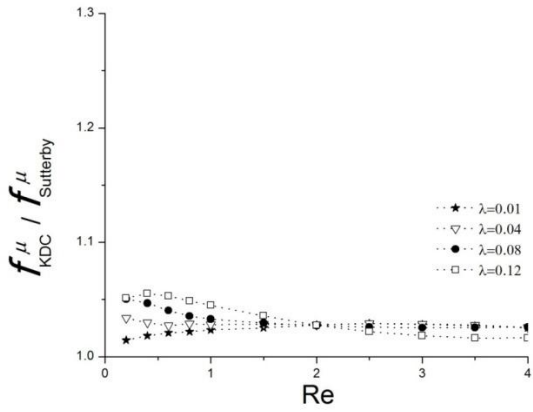
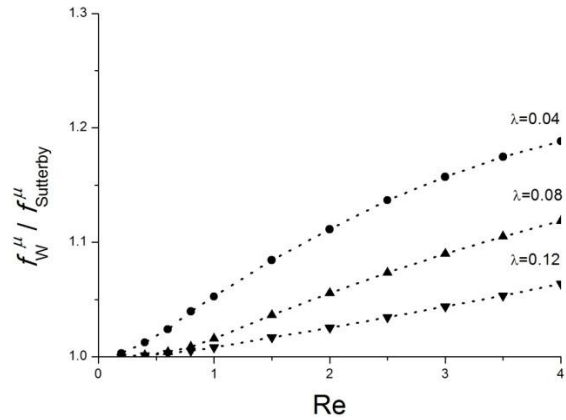

**Figure 7a:** Ratio of $f^\mu_{KDC}$ and Sutterby's $f^\mu$ as a function of $Re$ for different values $\lambda$ of specified in the figure. The range of $Re$ and $\lambda$ are those of Sutterby[38].

**Figure 7b:** Ratio of $f^\mu_W$ and those of Sutterby as a function of $Re$ for different values of $\lambda$ specified in the figure. The range of $Re$ and $\lambda$ are those of Sutterby [38].

**4.3.5.2 Wham and Sutterby:** In Fig. 7b we compare values of $f^\mu$ as given by simulation studies of Wham et.al[41] with those given in the correlation tables of Sutterby[38], derived from experiments. The agreement deteriorates with increasing value of $Re$ ($Re < 4$), for each $\lambda$. The deterioration is more rapid for $\lambda = 0.04$ which is outside the range of validity of the relation given by Wham et.al [41] .The agreement is inferior at $\lambda = 0.08$, the lower end of the validity range than at $\lambda = 0.12$, which is well within the range. Sutterby's table does not go beyond $\lambda = 0.13$. In the restricted range of $\lambda$ where both correlations hold, Wham equation starts showing significant deviation from Sutterby's results even at $Re$ as low as 2.

**4.3.5.3 KDC and Wham:** Fig. 8 shows the plot of the ratio $f^\mu_{KDC} / f^\mu_W$, calculated using respective equations, at several different values of $\lambda$ as a function of $Re_\infty$. The agreement is good at low values of $Re_\infty$ ($< 5$) for $\lambda \leq 0.4$.

Deviations from unity increase rapidly as $Re_\infty$ increases at larger $\lambda$ (0.4, 0.5), but remains small (<10%) at smaller $\lambda$ over the $Re_\infty$ in shown in Fig. 8.

At $\lambda = 0.3$ however, the ratio goes through a minimum, ~0.95 at $Re_\infty$~12 and then monotonically approaches to unity. The agreement between the predictions of the two equations, although 'good' at $\lambda = 0.1$ and 0.2, is best at (and around) $\lambda = 0.3$ over an extended $Re_\infty$ range. At $\lambda = 0.3$, $\mu_\infty^W$ is closest to $\mu_\infty^{actual}$, but $\mu_\infty^{KDC}$ differs more significantly from $\mu_\infty^{actual}$ (Table 4). In line with this $\Delta f_W^\mu/f^\mu$ is lowest(and smaller than $\Delta f_{KDC}^\mu/f^\mu$) at around $\lambda = 0.3$, where it switches sign, presumably going through zero. (Table 8). We infer Wham equation is more correct than KDC equation around $\lambda = 0.3$. Wham et. al. [41] compare the predictions of their simulation results with experimental results of McNown [37] at a value of $\lambda$ very close to this value ($\lambda = 0.3125$) and report good agreement in the $Re$ range $0 - 40$ (Fig. 14 of Wham et. al. [41]). In sec. 4.5 when $f_{KDc}^\mu$ is modified to bring $\mu_\infty^{KDC}$ at different values of $\lambda$, as a set, closer to $\mu_\infty^{actual}$, the values of $f_{KDC}^\mu$ in the modified form are greater than those of $f_{KDC}^\mu$ in its original form and come closer to $f_W^\mu$, at $\lambda = 0.3$ over an extended $Re_\infty$ range.

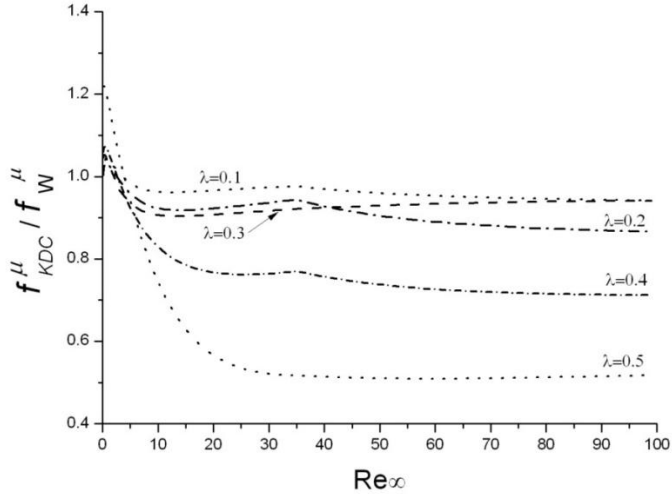

**Figure 8: Plot of the ratio of $f_{KDC}^\mu$ and $f_W^\mu$ versus $Re_\infty$.**

The values of $\mu_\infty^W$ (Table 4) as a set, differ more significantly from $\mu_\infty^{actual}$ than do the set $\mu_\infty^{KDC}$. The deviation increases more rapidly at a larger $\lambda$ for $\mu_\infty^W$. Deviation of $\mu_\infty^{KDC}$ from $\mu_\infty^{actual}$ never becomes as large as $\mu_\infty^{W\ values}$ at large $\lambda$ The same observations hold for the values of $Re_\infty$ given in Table 3.

In Fig. 9 we compare values of $f^\mu$ calculated from the experimental data points determined by K-D and reported in their paper (for details: sec. 4.3.2.2) and that calculated from Wham equation for two different values of $Re_\infty$ (0.01 and 10). The larger value of $\mu_\infty^{KDC}$ at $Re_\infty = 0.01$ is consistent with the result that in this range K-D

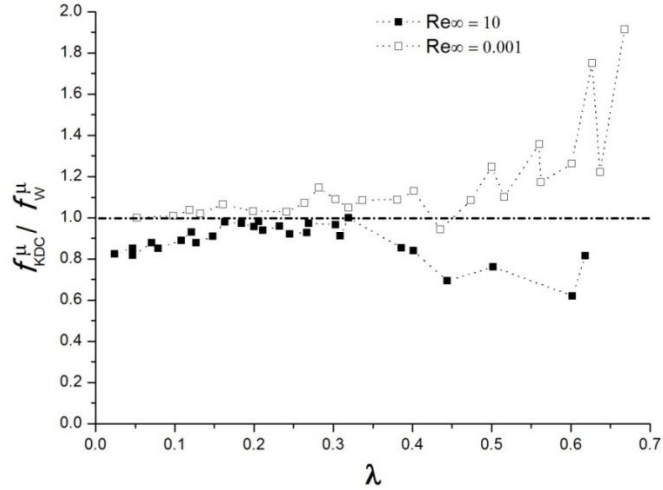

**Figure 9:** Plot of the ratio of $f^\mu_{KDC}$ and $f^\mu_W$ versus $\lambda$. Values of $f^V_{KD}$ are taken from Fig. 2 and 7 in K&D and used in calculation of $f^\mu_{KDC}$.

equation gives larger values of $f^V$ as compared to the 'exact theory' and Wham equation nearly reduces to H-S equation, which is very close to 'exact theory' (sec. 4.3.4, Table 10). The plot for $Re_\infty = 10$ shows agreement better than 10% at $\lambda \leq 0.3$, deteriorates significantly beyond $\lambda = 0.4$ and $f^\mu_W > f^\mu_{KDC}$ at all $\lambda$ This is consistent with Fig.8 and the larger values of $\mu_\infty^W$ as compared to $\mu_\infty^{actual}$ ($f^\mu_W > f^\mu_{exact}$) for $\lambda > 0.4$

We infer that KDC equation is overall superior to Wham equation in the range of $Re_\infty$ (~0.2 − 40) covered in experiments on Glycerol and Silicone oil. The values given by K-D equation at $Re < 1$ are slightly error-contaminated, while those given by Wham equation in this Re range is more accurate than is KDC around $\lambda = 0.3$.

**4.4 Test of the method at large $Re_\infty$: Data of Fidleris and Whitmore [36]:**

**4.4.1 $V_t$ data of Fidleris and Whitmore [36]:** The method has been tested as described in sec. 4.2 in the smaller $Re_\infty$ range (0.3 − 45). We now investigate its usefulness in a range of much larger $Re_\infty$ (~140 − 64000). Wham equation is not valid beyond $Re \sim 200$ and therefore, we use only KDC equation. We use the experimental terminal velocity data of 57 falls reported by F&W in their Fig. 1 for a wide range of $\lambda$ values in tubes of four different diameters on the test liquid water at 20 °C, a liquid having coefficient of viscosity ($= 0.001 Pa.s$) about three orders of magnitude smaller than that of Silicone oil or Glycerol. The values of $V_t$ in these measurements at large $Re_\infty$ are a factor of 5 or more, larger than in their low $Re_\infty$ counterparts (Table 2). Accuracy of the measurements as determined by F&W is 0.3%. Reading error increases this value marginally.

In Table 11 we give a subset of $V_t$ values given by F&W in their Fig. 1 for which, Eq. 13a could be solved. These values show a non monotonic dependence on $d$ at a fixed $D$ (a non monotonic $\lambda$ dependence), an increase of $V_t$ with increasing $d$, a maximum followed by a decrease. Numerical values (Table 11 and Fig.1 of F&W) are: $D = 15mm$, a maximum in $V_t$ (0.924m/s) is observed with $d = 6.315mm$, $\lambda = 0.421$; $D = 20mm$, $V_t^{max}$ (1.070m/s) with $d = 7.94mm$, $\lambda = 0.0397$; $D = 25mm$, $V_t^{max}$ (1.190m/s) with $d = 9.670mm$, $\lambda = 0.0380$; $D = 30mm$, $V_t^{max}$ (1.290m/s) with $d = 12.84mm$, $\lambda = 0.428$. The pattern in these data is a shift of $d$ for $V_t^{max}$ to larger values as $D$ increases, $\lambda$ remaining about the same. This pattern was observed with corresponding data at low $Re_\infty$ obtained with Glycerol and Silicone oil as test liquid, reported in sec. 4.2.1. The pattern of variation of $\mu_S$ and $Re_S$, as $d$ varies for a fixed $D$ (Table 11) is the same as for Glycerol and Silicone oil (Table 2).

**Table 10: Analysis of terminal velocity data of Fidleris and Whitmore for two tube diameter (20mm and 25mm) as a function of incerasing $\lambda$ by use of KDC equation. $\mu_\infty$ of water at 20°C is 0.00105 Pa.s (Fidlesris and Whitmore (1961))**

| | d (mm) | $\lambda$ | $V_t$(m/s) | $\mu_s$(Pa.s) | $Re_S$ | $Re_\infty^{V_t}$ | $Re_\infty^\mu$ | $\mu_\infty$(Pa.s) | $\Delta\mu_\infty/\mu_\infty$ |
|---|---|---|---|---|---|---|---|---|---|
| D = 20mm | 0.720 | 0.036 | 0.299 | 0.006 | 33.86 | 262.2 | 191.5 | 8.20E-04 | 2.19E-01 |
| | 1.154 | 0.058 | 0.423 | 0.012 | 42.28 | 489.0 | 461.3 | 1.00E-03 | 4.76E-02 |
| | 1.586 | 0.079 | 0.529 | 0.017 | 48.12 | 787.9 | 805.0 | 1.07E-03 | -1.90E-02 |
| | 2.020 | 0.101 | 0.621 | 0.024 | 52.06 | 1194.3 | 1206.8 | 1.06E-03 (-3.00E-05) | -9.52E-03 |
| | 2.400 | 0.120 | 0.689 | 0.031 | 53.94 | 1596.4 | 1596.4 | 1.05E-03 (-8.00E-05) | 0.00E+00 |
| | 2.820 | 0.141 | 0.764 | 0.038 | 56.44 | 3543.9 | 2062.0 | 6.20E-04 (-3.80E-04) | 4.10E-01 |
| D = 25mm | 0.600 | 0.024 | 0.303 | 0.004 | 41.73 | 466.5 | 134.2 | 3.90E-04 | 6.29E-01 |
| | 1.173 | 0.047 | 0.426 | 0.012 | 42.21 | 480.1 | 474.7 | 1.04E-03 | 9.52E-03 |
| | 1.565 | 0.063 | 0.524 | 0.017 | 47.84 | 754.0 | 786.9 | 1.09E-03 | -3.81E-02 |
| | 1.988 | 0.080 | 0.625 | 0.023 | 53.60 | 1341.5 | 1175.0 | 9.30E-04 (-4.00E-05) | 1.14E-01 |
| | 2.350 | 0.094 | 0.689 | 0.029 | 55.09 | 1698.3 | 1543.3 | 9.60E-04 (-9.00E-05) | 8.57E-02 |
| | 2.825 | 0.113 | 0.770 | 0.038 | 57.23 | 3141.9 | 2067.7 | 7.00E-04 (-3.30E-04) | 3.33E-01 |

**4.4.2 $Re_\infty^\mu, Re_\infty^{V_t}, \mu_\infty$:** The values of $Re_\infty^\mu$ and $Re_\infty^{V_t}$ are calculated by methods specified in sec. 3.4.1 and 3.4.2 respectively These values, along with those of $\mu_S$, $Re_S$, $\mu_\infty$ and its relative deviation $\Delta\mu_\infty/\mu_\infty$, in those cases where a solution was obtained, are given in Table 11. The inequality $Re_\infty^\mu \neq Re_\infty^{V_t}$ noted in Table 3 (sec. 4.2.6) for low $Re_\infty$ systems is observed. The values of $Re_\infty^\mu$ are larger, so are the differences between $Re_\infty^\mu$ and $Re_\infty^{V_t}$, compared to those in Table 3. $Re_\infty^\mu > Re_\infty^{V_t}$ in all entries of Table 3. In contrast, in Table 11 the difference is not monotonic and changes sign as $\lambda$ increases. The values of $\mu_S$ and $Re_S$ show trends observed in sec. 4.2.5 for low $Re_\infty$ systems.

We also find in Table 11 that $\mu_\infty$ of water vary with change in $d, D$. This feature and the inequality $Re_\infty^\mu \neq Re_\infty^{V_t}$ indicate an error. We assign, as we did for low viscosity results, an inexact functional form of the correlation equations as the source of this error. In calculation of $\mu_\infty$ from $V_t$, we obtain unique values only for very low values of $\lambda$. With increasing values of $\lambda$, a pair of values of $\mu_\infty$ are observed. It is apparent from Fig. 4 that the two values come closer with increasing $Re_S$, as they do in Table 11. The values of $Re_\infty^{V_t}$ in the largest $\lambda$ entry for both tube diameters in Table 11 approach the value at the maximum in Fig. 4. In this sequence, for yet larger values of $Re_S$, no root was obtained in the whole range of scanned $Re_\infty$. One of the two values of $\mu_\infty$, in systems that give a root pair, is very close to the actual value. The other value is given in bracket in Table 11. $\lambda$ being low, the accuracy of the unique roots and the more 'correct' root where a root pair is present, are excellent and deteriorates only slightly in the largest $\lambda$ entry of both tube diameters. In Fig. 11(a-d) we show values of $\mu_\infty$, a single value or a pair of values as the case may be, for all four tubes and discuss their accuracy in sec. 4.4.6.

**Table 12: Values of calculated and 'exact' velocity and viscosity ratio for high $Re_\infty$**

| | $\lambda$ | $f^V(\mu, V_t)$ | $f^V_{KD}$ | $f^\mu(\mu, V_t)$ | $f^\mu_{KDC}$ |
|---|---|---|---|---|---|
| D = 20mm | 0.036 | 1.054 | 0.999 | 0.158 | 0.129 |
| | 0.058 | 0.999 | 0.997 | 0.087 | 0.087 |
| | 0.079 | 0.985 | 0.994 | 0.058 | 0.062 |
| | 0.101 | 0.986 | 0.990 | 0.042 | 0.044 |
| | 0.120 | 0.981 | 0.985 | 0.033 | 0.034 |
| | 0.141 | 0.991 | 0.979 | 0.026 | 0.016 |
| D = 25mm | 0.024 | 1.270 | 1.000 | 0.231 | 0.090 |
| | 0.047 | 0.993 | 0.998 | 0.085 | 0.088 |
| | 0.063 | 0.985 | 0.997 | 0.059 | 0.064 |
| | 0.080 | 1.002 | 0.994 | 0.044 | 0.040 |
| | 0.094 | 0.995 | 0.991 | 0.034 | 0.033 |
| | 0.113 | 0.999 | 0.987 | 0.027 | 0.018 |

**4.4.3 $f^V$ and $f^\mu$:** In Table 12 we give values of $f^V(\mu, V_t), f^V_{KD}, f^\mu(\mu, V_t), f^\mu_{KDC}$, for the falls analysed in Table 11 We note the following features in $f^V$ and $f^\mu$ of Table 12: (i) Pattern of variation of $f^V$ and $f^\mu$ with change in $\lambda$ and $Re_\infty$ is the same as in low $Re_\infty$ system (Table 5, 6). (ii)The values of $f^V_{KD}$ at larger $Re_\infty$ are larger(~1) as compared to their counterparts (similar $\lambda$) at smaller $Re_\infty$ (Table 5). This is consistent with experimental results cited by Kehlenbeck and DiFelice [34]. This feature of larger $f^V$ values at larger $Re_\infty$ is also observed for values of $f^V(\mu, V_t)$. (iii) In several cases $f^V(\mu, V_t)$ exceeds unity. This indicates an error. We return to this issue in sec. 4.4.8. (iv) The values of $f^\mu$ in large $Re_\infty$ systems (Table 12) are orders of magnitude smaller than their counterparts (similar $\lambda$) in systems with lower $Re_\infty$, given in Table 5. This is a consequence of much larger values of $(F_D/F_S)_{\lambda \to 0}$ (Eq. 12) for larger $Re_\infty$ ($\sim 200 - 2000$) as compared to smaller $Re_\infty$ ($\sim 0.2 - 40$) ball-liquid systems; an increase in $Re_\infty$ by 2 orders of magnitude from 20 to 2000 increases $(F_D/F_S)_{\lambda \to 0}$ from 2.27 to 35.05. A smaller $f^\mu$ implies a larger wall-correction.

**4.4.4 $\Delta f^V$, $\Delta f^\mu$:** Except for two anomalous ($f^V > 1$) lowest $\lambda$ entries (sec. 4.4.8), $\Delta f^V$ and $\Delta f^\mu$ are small, particularly because the solutions could be obtained only at small $\lambda$. In Table 13 we give details of $\Delta f^V \to \Delta f^\mu$ translation (sec. 3.4.7.2), the values of $\Delta f_1^V/f^V$, $\Delta(F_D/F_S)_{\lambda \to 0}/(F_D/F_S)_{\lambda \to 0}^{Re_\infty^\mu}$, $\Delta f_1^\mu/f^\mu$, $\Delta f_2^V/f^V$ and $\Delta f_2^\mu/f^\mu$. As noted in sec. 3.4.7.2 in these large $Re_\infty$ systems, an amplification as well as a change in sign of $(\Delta f_1^\mu/f^\mu)$ with respect to $(\Delta f_1^V/f^V)$ is possible, depending on $\Delta(F_D/F_S)_{\lambda \to 0}/(F_D/F_S)_{\lambda \to 0}^{Re_\infty^\mu}$. Both of these features are observed in Table 13. Apart from the anomalous lowest $\lambda$ entries (sec. 4.4.8) $\Delta f_1^\mu$ is a more significant contributor to $\Delta f^\mu$ than is $\Delta f_2^\mu$. This is in contrast to low $Re_\infty$ systems (Table 8). This is most clearly seen in the largest $\lambda$ entry in both tables (0.369 vs. 0.013; 0.292 vs. 0.012).

**Table 11: Translation of $\frac{\Delta f_1^V}{f^V} \to \frac{\Delta f_1^\mu}{f^\mu}$ and $\frac{\Delta f_2^V}{f^V} \to \frac{\Delta f_2^\mu}{f^\mu}$ in high $Re_\infty$ cases.**

|  | $\lambda$ | $\frac{\Delta f_1^V}{f^V}$ | $\frac{\Delta(F_D/F_S)_{\lambda \to 0}}{(F_D/F_S)_{\lambda \to 0}}$ | $\frac{\Delta f_1^\mu}{f^\mu}$ | $\frac{\Delta f_2^V}{f^V} = \frac{\Delta f_2^\mu}{f^\mu}$ |
|---|---|---|---|---|---|
| D=20mm | 0.036 | 0.000 | -0.160 | 0.131 | 0.052 |
|  | 0.058 | 0.000 | -0.004 | 0.004 | 0.002 |
|  | 0.079 | 0.000 | 0.053 | -0.056 | -0.009 |
|  | 0.101 | 0.000 | 0.050 | -0.053 | -0.004 |
|  | 0.120 | 0.000 | 0.038 | -0.040 | -0.003 |
|  | 0.141 | -0.001 | -0.599 | 0.369 | 0.013 |
| D=25mm | 0.024 | 0.000 | -1.022 | 0.398 | 0.212 |
|  | 0.047 | 0.000 | 0.027 | -0.028 | -0.006 |
|  | 0.063 | 0.000 | 0.067 | -0.073 | -0.012 |
|  | 0.080 | 0.000 | -0.072 | 0.067 | 0.008 |
|  | 0.094 | 0.000 | -0.045 | 0.043 | 0.004 |
|  | 0.113 | 0.000 | -0.419 | 0.292 | 0.012 |

$\Delta f_2^V$ in larger $Re_\infty$ cases is of comparable magnitude to its counterparts for smaller $Re_\infty$ cases. This is to be anticipated since these are purely form errors arising from defect in functional form alone (sec. 3.4.8.3) and may not show any specific pattern in their dependence on $(\lambda, Re_\infty)$. However, in the large $Re_\infty$ systems $\Delta f_2^\mu$ is reduced to a much smaller value as compared to its small $Re_\infty$ counterparts (Eq. 24e; much larger $(F_D/F_S)_{\lambda \to 0}^{Re_\infty^\mu}$). The ratios $\Delta f_2^V/f^V$ and $\Delta f_2^\mu/f^\mu$ are however equal (Table 13).

$\Delta f_1^V$ is smaller in water than their counterparts in smaller $Re_\infty$ systems because of insensitivity of $f_1^V$ to change in $Re_\infty$ as $Re_\infty$ grows larger. Even though the magnitude of $(Re_\infty^\mu - Re_\infty^{V_t})$ is significantly larger for water as compared to Glycerol and Silicone oil, the sensitivity of $f_{KD}^V$ towards $Re_\infty$ is so low at larger $Re_\infty$ that $\Delta f_1^V$ in water is negligibly small. $f^V$ is, however of the same order in water as compared to Glycerol/Silicone oil. (Table 9 and 13); As a result, the transformation of $\Delta f_1^V$ to $\Delta f_1^\mu$ is determined almost wholly by values of $(F_D/F_S)_{\lambda \to 0}$ as given in Eq. 25.

**4.4.5 $\Delta f^\mu$ and $\Delta f^\mu/f^\mu$:** In large $Re_\infty$ systems $\Delta f^\mu$ and $f^\mu$ are both orders of magnitude smaller for water as compared to Glycerol/Silicone oil but $\Delta f^\mu$ may not have decreased by the same factor as does $f^\mu$. As a result $\Delta f^\mu/f^\mu$ may assume larger values (Table 9 and 13), the largest in the set under study being $\sim 2.8$ (largest $\lambda$ system of $D = 30mm$ set).

**4.4.6 Calculated $\mu_\infty$, $Re_\infty^{V_t}$: Uniqueness and accuracy:** In 12/57 falls (all 4 tube diameters), for 11 of which $\lambda$ is small (below $\sim 0.1$), an unique value of $\mu_\infty$ (and $Re_\infty^{V_t}$) could be calculated. These 11 systems are easily identified in Fig. 11(a-d): 3 systems each in Fig. 11(a), 11(b) 11(c) and 11(d). The numerical value of $\mu_\infty$ for each solution is given in the figures and for $D = 20$ and $25mm$ also in Table 11. Five of these are accurate to $\sim 10\%$, three are in the range $10 - 20\%$, three others have larger errors, namely $\sim 62\%$ (the lowest $\lambda$ system in Fig.11(c)), and 30 and 34% (the two lowest $\lambda$ systems Fig. 11(d)). Values of $Re_S$ (and $\lambda$) for these unique solutions are smaller than those of systems with multiple roots, as is expected (Fig. 4) The twelfth unique solution was obtained for $D = 15mm$, $\lambda = 0.842$, a larger $\lambda$, accompanied by a smaller $Re_S$, $2.37$ compared to those for multiple root cases, $51.99, 53.10$ and $54.55$ for the three multiple root cases in order of increasing $\lambda$ (non-monotonic variation of $Re_S$ with $\lambda$). This result is consistent with Fig. 4. Calculated $\mu_\infty$ is $2.92 \times 10^{-3}$, in error by $\sim -180\%$. We have earlier noted that the form of K-D equation is error-prone at large $\lambda$.

In 12/57 falls for which $\lambda$ is intermediate ($\sim 0.1 < \lambda < \sim 0.2, \sim 0.3$ in $D = 30mm$), a pair of values of $\mu_\infty$ (and $Re_\infty^{V_t}$) are found, 3 each in Fig. 11(a-d). Of these three systems in each figure the two with lower $\lambda$ have one root very close to the correct value of $\mu_\infty$; the other root is not so close. A larger deviation of both roots from the 'correct root' is found for the largest $\lambda$ system. In $D = 15, 20, 25mm$, the difference between the two solutions decreases with increasing $Re_S$ as is expected from Fig. 4. The choice between the two values is then ambiguous even if there is an approximate idea of the 'correct' value. The 'more correct' root is accompanied by a $Re_\infty^{V_t}$ close to $Re_\infty^\mu$. An example: In a multiple root case, the two pairs of $(Re_\infty^{V_t}, \mu_\infty)$ are $(4.21 \times 10^4, 3.00 \times 10^{-4})$ and $(1.19 \times 10^3, 1.06 \times 10^{-3})$; the corresponding $Re_\infty^\mu$ is $1.27 \times 10^3$, very close to $1.19 \times 10^3$ ($\mu_\infty$ values are also very close, $1.05 \times 10^{-3}$ and $1.06 \times 10^{-3}$ respectively) and an order of magnitude different from $4.21 \times 10^4$. The 6 systems with a pair of roots for $D = 20$ and $25 mm$ are given in Table 11. In $D = 30 mm$ the separation between the two roots increases to a small extent in the largest $\lambda$ system compared to the other two multiple root cases. This order corresponds to the values of $Re_S$ for the three multiple root systems, $54.46, 56.32$ and $44.11$ in increasing order of $\lambda$. In 33 cases with yet larger $\lambda$ ($> 0.3$) and $Re_S$, values of $\mu_\infty$ (and $Re_\infty^{V_t}$) could not be obtained.

We infer from these results that KDC equation gives satisfactory results with low viscosity liquids, just as it does for highly viscous liquids. In the range of $\lambda$ investigated $Re_\infty$ increases with increasing $\lambda$ (Table 11). The problems encountered with increasing $\lambda$ referred to above arise from increasing $Re_\infty$. This issue has already been discussed (sec. 3.4.2 and 3.4.5, Fig. 4).

**4.4.7** $\mu_\infty \to V_t, Re_\infty^\mu, f^V(\mu), f^\mu(\mu)$ : **KDC calculations:** In another set of calculation, we use the known value of $\mu_\infty$ of water ($0.001 Pa.s$) to calculate $Re_\infty^\mu$, $f^V(\mu)$, $f^\mu(\mu)$ and $V_t$ (sec. 3.4.1). With known values of $\mu_\infty$ and $V_t$ we calculate $f^V(\mu, V_t)$ and $f^\mu(\mu, V_t)$ (sec. 3.4.6) and then $\Delta f^V, \Delta f^\mu, \Delta f_{1,2}^V, \Delta f_{1,2}^\mu$. In Fig. 10 we give plots of $V_t$ versus $\lambda$ for 4 different tubes for the whole set of falls reported by F&W.

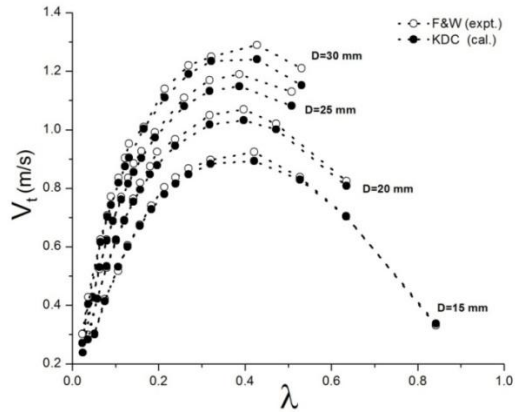

**Figure 8:** Plot of experimental values of $V_t$ vs $\lambda$ as given in Fig. 1 of F&W; and those calculated in this paper $\mu_\infty = 0.00105$ Pa.s by method given in sec. 3.4.1.

The agreement of values of $V_t$ calculated from known value of $\mu_\infty$ of water ( sec. 3.4.1) with experimental values reported by F&W is excellent (Fig. 10). The maximum deviation is $< 5\%$. In sec. 4.3.2.1 we estimate the accuracy of F&W and find that ~5% is a reasonable estimate. The excellent accuracy and that it is possible to obtain values of $V_t$ in all cases is a result of a more accurate $Re_\infty$, namely $Re_\infty^\mu$ being available, as against $Re_\infty^{V_t}$ in $V_t \to \mu_\infty$ calculations. This calculation assumes importance in view of the fact that in many applications the focus may be on calculation of terminal velocity in a liquid of known viscosity coefficient in a given tube and not on determination of viscosity coefficient of an unknown liquid.

**4.4.8 $f^V > 1$**: In several ball-tube combinations we have obtained $f^V(\mu, V_t) > 1$, an anomalous result. There are 3 such values in Table 12, viz., $1.054, 1.270$ and $1.002$. Similar anomalous systems have been observed in other tube diameters, viz., $D = 15mm$, first 2 falls (lowest $\lambda$ and $V_t$), ($\lambda = 0.051$, $f^V = 1.01$) and $(0.113, 1.02)$; $D = 30mm$, first 4 falls, $(0.023, 1.11)$, $(0.036, 1.06)$, $(0.065, 1.01)$ and $(0.081, 1.01)$. An error is implied. Its calculation uses the value of $V_t$ and Cheng equation (Eq. 16). Use of Clift equation marginally alters the values of $f^V$ and do not remove the anomaly. The scatter of experimental values of drag ratio around those predicted by Clift equation (sec. 4.3.1) also contributes to this anomaly. Even after these errors are considered, $1.27$ is indeed an outlier. We note that the four cases with large $f^V$ anomaly ($> 1.05$) appear for systems with low $Re_\infty$ ($142 - 436$).

**4.4.9 Anomalous $Re_S$**: We have noted that over the whole range of $Re_\infty$, $Re_S/(f^V)^2$ ($\lambda$ independent) assumes a maximum value of 59.07 at $Re_\infty = 4363$. For a given $\lambda$, $f^V$ assumes a maximum value at a large $Re_\infty \sim 10^4$. This maximum value increases with decreasing $\lambda$ (Fig. 3 and 4 of F&W). Consequently, $Re_S \left(= \frac{Re_\infty}{(F_D/F_S)_{\lambda \to 0}} (f^V)^2\right)$ has a maximum value over the whole range of $Re_\infty$, for a given $\lambda$ which increases with decreasing $\lambda$ (Table 14). In Table 14, for the systems with $\lambda \geq 0.157$ the experimental value of $Re_S$ reported by F&W exceeds the calculated maximum $Re_S$. The maximum values of $Re_S$ cited in Table 14 depend on the value of $f^V$ used and in these calculations we have used $f^V_{KD}$. Modification of the K-D formula that we consider in sec. 4.5 increase the value of $f^V$, lower the maximum limit on $Re_S$ and the anomaly is removed. The anomaly arises from an incorrect estimate of $f^V$ as given by K-D formula. Scatter around the predictions of Cheng equation will also contribute to this anomaly. The values for the scatter around Clift equation predictions are in sec. 4.3.1.

Table 12: Maximum possible value of $Re_S$ for $\lambda$ values given in Table 13.

| $\lambda$ | $Re_S$ (Max) | $Re_S$(F&W) |
|---|---|---|
| 0.036 | 58.96 | 33.86 |
| 0.058 | 58.75 | 42.28 |
| 0.079 | 58.42 | 48.12 |
| 0.101 | 57.92 | 52.06 |
| 0.120 | 57.36 | 53.94 |
| 0.141 | 56.60 | 56.44 |
| 0.157 | 55.92 | 58.25 |
| 0.180 | 54.79 | 58.00 |
| 0.196 | 53.89 | 59.52 |
| 0.238 | 51.14 | 53.57 |
| 0.318 | 44.52 | 47.27 |
| 0.397 | 36.73 | 39.32 |
| 0.472 | 28.93 | 30.05 |
| 0.635 | 13.62 | 14.58 |

**4.5 Results with modified KDC equation:** We incorporate a modification in KDC equation by adding a forth order polynomial in $\lambda$ with constant term set to be zero, to the $f^V$ component of KDC expression of $f^\mu$. The coefficients of the polynomial are chosen to minimize the RMS difference between calculated values of a $\mu_\infty$ set at different $Re_\infty$ and $\lambda$ and the 'correct' $\mu_\infty$. We find that the coefficients depend on the range of $Re_\infty$ and $\lambda$ chosen. The coefficients are therefore functions of both $Re_\infty$ and $\lambda$.

**4.5.1 Large $Re_\infty$ systems**: Results with modified KDC equation show considerable improvements over those obtained with KDC equation in its original form. Of the 33 ball-tube combinations where no solution was obtained with original KDC equation, 12 give solutions with modified KDC equations. In these 12 new solutions the value of $\lambda$ (and $Re_S$) is larger than the values of $\lambda$, for which KDC solutions were obtained, with only one exception, largest $\lambda$ solution in $D = 30mm$ is a KDC solution. With the 'old' functional form of the K-D equation values of $f^V$ were such that no point of intersection was obtained in Fig.4; the modification of the functional form alters the value of $Re_S/(f^V)^2$ vs. $Re_\infty$ such that its plot shifts to smaller values on the y-axis in Fig. 4 and a point of intersection is obtained. These new solutions are identified in Fig. 11(e-h). The distribution is as follows: 3 in Fig. 11(e), 3 in Fig. 11(f), 4 in Fig. 11(g), 2 in Fig 11(h). The numerical value of $\mu_\infty$ of each solution, unique or multiple is indicated in the figures. We summarize the main features of the solutions, in particular their accuracy, separately for solutions obtained from KDC equation in its original form and those from its modified version.

The accuracy of $\mu_\infty$ in all of 11 high accuracy low $\lambda$ unique KDC solutions marginally alter when modified KDC equation is used. The separation between the two values of $\mu_\infty$ in all of 12 ball-liquid pairs with multiple roots in the original KDC equation, increase with modified KDC. This is expected in view of the shift of the plot of $Re_S/(f^V)^2$ vs. $Re_\infty$ to smaller values on the y-axis of Fig.4 as a result of the modification. One of the roots in each of the 12 solutions obtained with modified KDC equation is close to the correct $\mu_\infty$. In the system triplets with multiple solutions in each figure, two lower $\lambda$ ball-tube combinations had one KDC root very close to the correct $\mu_\infty$. The accuracy of this root marginally alters in modified KDC solutions. In tubes with $D = 15, 20, 25 mm$, the ball-tube combination with largest $\lambda$ amongst the triplet gives two roots, both smaller than the correct $\mu_\infty$. In the modified KDC solutions, the separation between the roots increases the larger value and brings it up close to the correct $\mu_\infty$. In largest $\lambda$ system of $D = 30mm$, the smaller of the two KDC $\mu_\infty$ values decreases in value, from larger than the correct value to smaller than it, relative error remaining nearly the same.

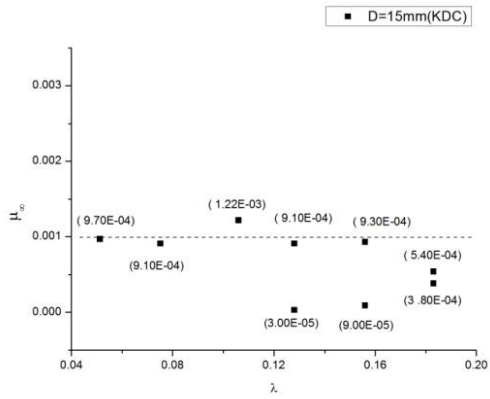

Figure 11a: D=15mm KDC

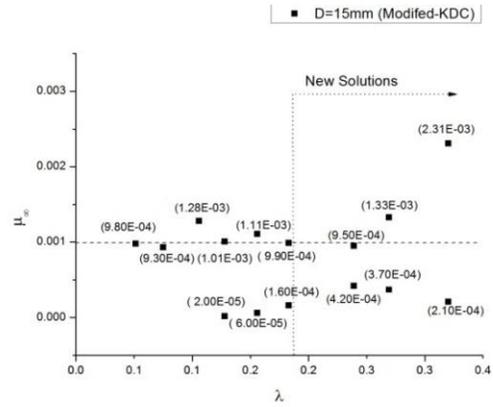

Figure 11e: D=15mm Modified-KDC

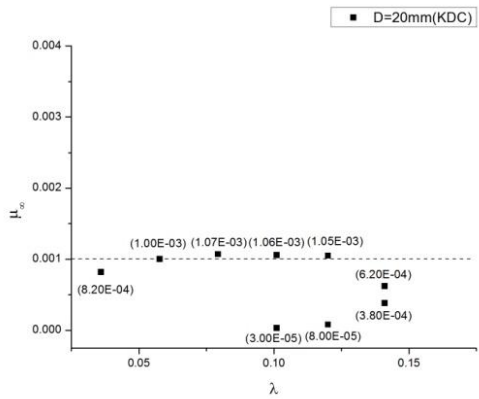

Figure 11b: D=20mm KDC

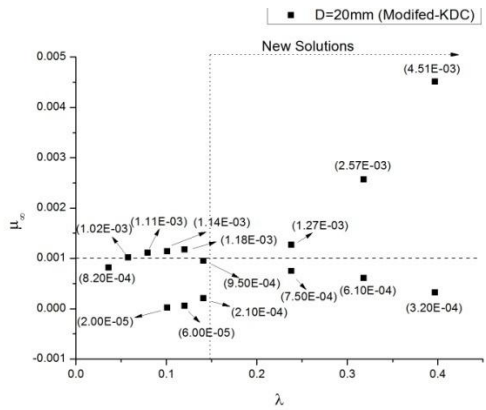

Figure 11f: D=20mm Modified-KDC

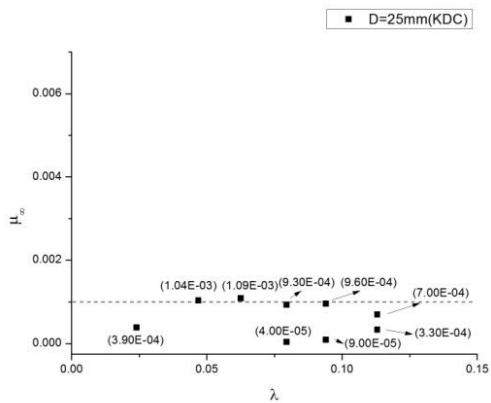

Figure 11c: D=25mm KDC

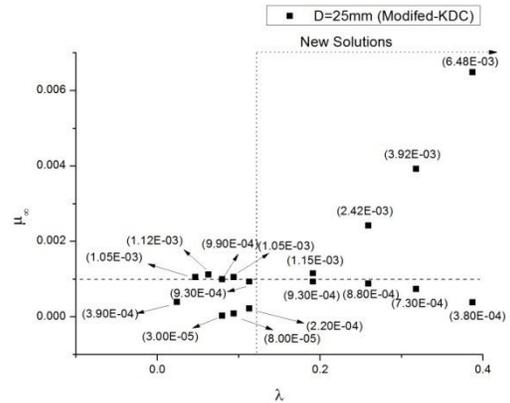

Figure 11g: D=25mm Modified-KDC

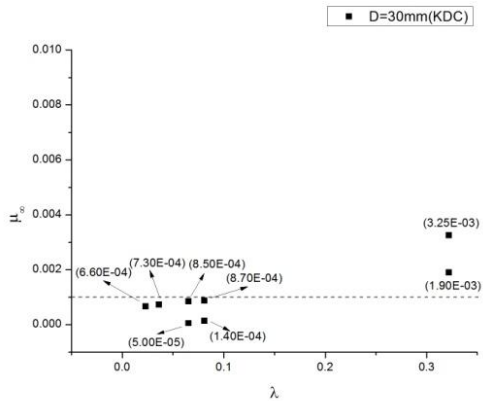
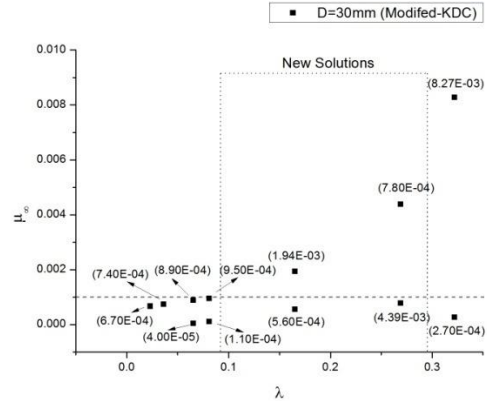

**Figure 11d: D=30mm KDC**  **Figure 11h: D=30mm Modified-KDC**

**Fig. 11 Values of $\mu_\infty$ calculated by KDC equation and its modified form from $V_t$ data of F&W in tubes of four different diameters.**

Each of the 'new' solutions appears as a pair. As $\lambda$ increases the separation in value between the pairs increases for each tube. This is a consequence of the fact that the modification $f^V$ is larger for larger $\lambda$ causing a larger shift to lower values on the y-axis of Fig.4, thereby increasing the separation between the two roots. The lowest $\lambda$ 'new' solution for $D = 15mm$, has one of the two values close to the correct value (~10%); for $D = 20, 25, 30mm$ the lowest $\lambda$ solution pair are equally spaced around correct $\mu_\infty$ with good accuracy (~10% in Fig. 11(g), ~25% in Fig. 11(f), ~40% and ~90% in Fig. 11(h)) Each of the 8 solutions at higher values of $\lambda$ have one of the two solutions close to the correct value, accuracy range being 25-80%. One of the two solutions in each case is within at most a factor of 2 of the correct $\mu_\infty$.

Modified KDC equation increases the range of $\lambda$ and $Re_\infty$ for which solutions of good accuracy can be obtained. RRMSD with respect to 'correct' value are: KDC: 0.465 and modified KDC: 0.322.

The values of the coefficients of the forth order polynomial in order of decreasing power of $\lambda$ are: [0.153338, -0.322906, 0.294622, 0.042879] for $\lambda < 0.4$ and [0.182216 0.267440 -0.661362 0.252113] for $\lambda > 0.4$.

**4.5.2 Low $Re_\infty$ systems:** Modified KDC equation improves the constancy of values of $\mu_\infty$ determined with different $\lambda$ and $Re_\infty$ in experiments with Glycerol and Silicone oil. In these low $Re_\infty$ systems the issue of multiple solutions or of new solutions with modified KDC equation does not arise. The results are shown for Glycerol in Fig. 12(a) (KDC) (Table 4) and in Fig. 12(b) (modified-KDC); for Silicone oil in Fig. 12(c) (KDC) (Table 4) and in Fig. 12(d) (modified-KDC). The RRMSD with respect to the correct value of $\mu_\infty$ decreases from 0.214 to 0.082 for Glycerol and from 0.0463 to 0.0262 for Silicone oil. Numerical values are inserted in the figures. We note that the figures for Glycerol are shown in a more expanded scale. The largest deviation in $\mu_\infty$ observed, in modified-KDC, for Glycerol with $\lambda = 0.436$ is 12.59% and that for Silicone oil with $\lambda = 0.667$ ($D = 16.5 mm$) is 42.16%.

The values of the coefficients of the forth order polynomial in order of decreasing power of $\lambda$ are: [-0.874601, -1.533334, 1.249985, -0.115640] for $\lambda < 0.4$ and [1.704933, -1.917485, 0.150617, 0.215884] for $\lambda > 0.4$.

In Fig. 13 we show a comparison among plots of $f_{KD}^V$ for different values of $\lambda$ for the whole range of $Re_\infty$ and $f_{KD}^V$ (modified) which is a component of modified-KDC. The modification is 'small', < 5% which lies well within the uncertainty of KD equation (sec. 4.3.2.2).

**4.6 Error in $\mu_\infty$:** Deviation of calculated $\mu_\infty$ from its correct value arises from (i) translation of error in $V_t$ (difference between mean $V_t$ and 'correct' $V_t$) to that of $\mu_\infty$ ($V_t$ has no systematic error) and from (ii) inexact functional form of correlation equations.

**4.6.1 $\Delta V_t \rightarrow \Delta \mu_\infty$:** Error in $V_t$ ($\Delta V_t$) translates into error of $Re_\infty$ which in turn gets translated into error of $\mu_\infty$. We use Eq. 11 to obtain

$$\frac{\Delta \mu_\infty}{\mu_\infty} = \frac{\Delta \mu_S}{\mu_S} + \frac{\Delta f^\mu}{f^\mu} \qquad (31)$$

$\Delta \mu_S$ arises from $\Delta V_t$; $\frac{\Delta \mu_S}{\mu_S} = \frac{\Delta V_t}{V_t}$. Using Eq. 31, the relation between $\Delta \mu_S$ and $\Delta V_t$ and Eq. 12 we obtain

$$\frac{\Delta \mu_\infty}{\mu_\infty} = 1 - \left(1 + \frac{\Delta V_t}{V_t}\right)\left(1 - \frac{\Delta f^V}{f^V}\right)\left(\frac{1}{1 - \frac{\Delta (F_D/F_S)_{\lambda \rightarrow 0}}{(F_D/F_S)_{\lambda \rightarrow 0}}}\right) \qquad (32)$$

It follows that the translation of $\Delta V_t/V_t \rightarrow \Delta \mu_\infty/\mu_\infty$ depends on the value of $Re_\infty$. $\Delta V_t$ is related to $\Delta Re_S$ by $\frac{\Delta Re_S}{Re_S} = 2\frac{\Delta V_t}{V_t}$ (Eq. 3, Eq. 5); as a result $Re_S/(f^V)^2$ is altered for a given functional form of $f^V(Re_\infty, \lambda)$, so do points of intersection and the two coordinates $\frac{Re_\infty}{(F_D/F_S)_{\lambda \rightarrow 0}}$ ($= \alpha$), $Re_\infty$ in Fig. 4. Values of $f^V$ (and then $\mu_\infty$) are modified according to Eq.33

$$\frac{\Delta \alpha}{\alpha} = \frac{\Delta Re_S}{Re_S} - \frac{2\Delta f^V}{f^V} \qquad (33)$$

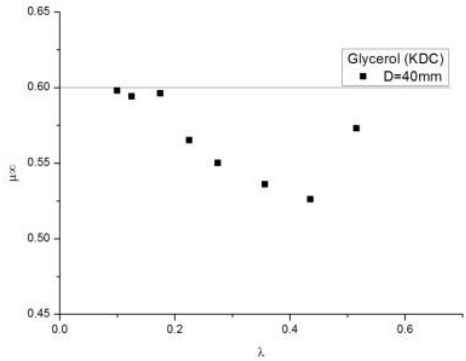 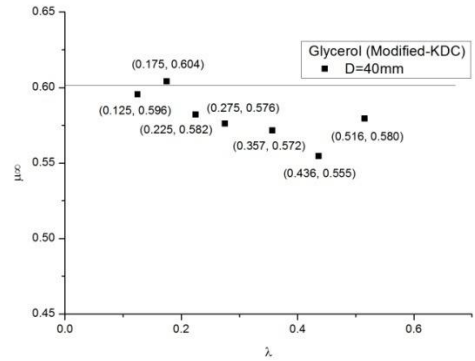

**Figure 12a: D=40mm; Glycerol; KDC**     **Figure 12c: D=40mm; Glycerol; Modified-KDC**

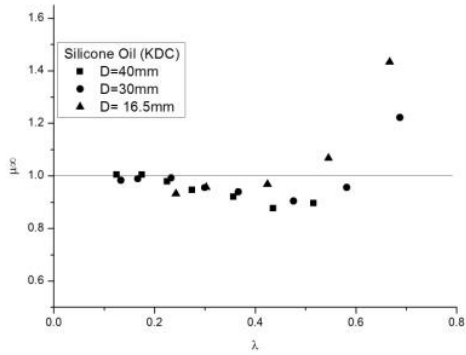 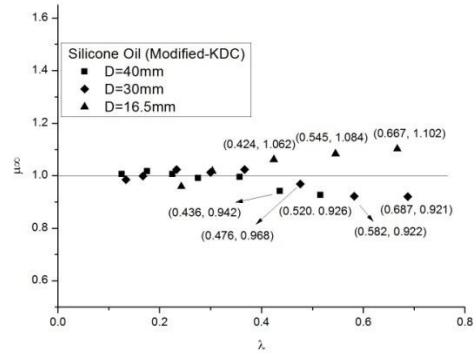

**Figure 12b: D=40, 30 and 16.5mm; Silicone oil; KDC**     **Figure 12d: D=40, 30 and 16.5mm; Silicone oil; Modified-KDC**

**Fig. 12: Values of $\mu_\infty$ calculated by KDC equation and its modified form for Glycerol & Silicone oil. $V_t$ data are in Table 2. $\mu_\infty^{KDC}$ values are in Table 4.**

We note that $\Delta f^V$ of Eq. 33, arising entirely from $\Delta Re_\infty$ is $\Delta f_1^V$ defined in Eq. 23a, has negligible value and can be ignored in Eq. 32 and 33. Table 15 and Fig. 14 shows propagation of error in $V_t$ to that in $\mu_\infty$ in different ranges of $Re_\infty$ and $\alpha$. The error is large for larger $Re_\infty$. The values stress the need to make highly accurate measurements of $V_t$ with large $Re_\infty$ ball-liquid combinations. The 22 ball-tube combinations for which no solutions could be obtained have large $Re_\infty$ and belong to the error-prone group.

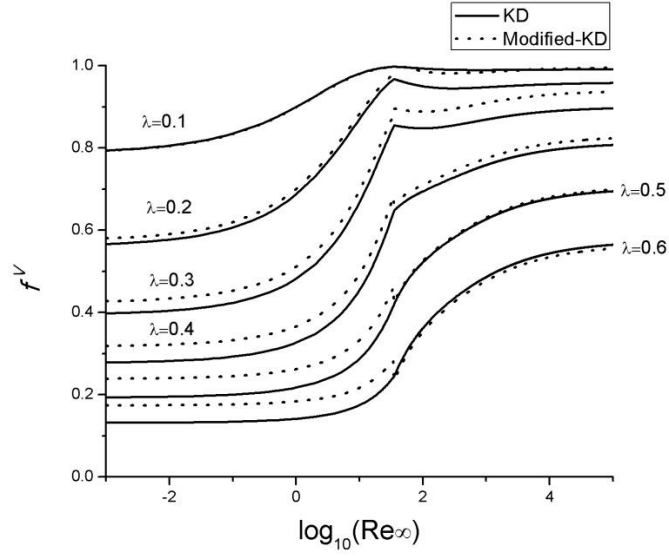

Figure 13: $f^V_{KD}$ for fixed value of $\lambda$ as a function of $Re_\infty$: original and modified form.

With $\Delta V_t = 0$ we obtain (using Eq. 25)

$$\frac{\Delta \mu_\infty}{\mu_\infty} = \frac{\Delta f_1^\mu}{f^\mu} \qquad (34)$$

**4.6.2 Functional form error**: The presence of error in the functional form of correlation equations generates a contribution to $\Delta f_1^\mu$ and a non zero $\Delta f_2^\mu$. We obtain

$$\frac{\Delta \mu_\infty}{\mu_\infty} = \frac{\Delta f^\mu}{f^\mu} \qquad (35)$$

**4.6.3 Error magnitudes**: An increase in $\Delta f^\mu / f^\mu$ and $\Delta \mu_\infty / \mu_\infty$ with increasing $\lambda$ and consequent increase in $Re_\infty$ is observed for low $Re_\infty$ systems (sec. 4.2.9, Table 8). A value of $\lambda > 0.4$ is found to be error-prone. This error is entirely functional form error. In large $Re_\infty$ domain we again find larger relative deviation in calculated $\mu_\infty$ (~33%, ~40%) for the largest $\lambda$ entry in $D = 20, 25 mm$ (Table 11), for which $Re_\infty$ values are also large ($Re_\infty^\mu$ ~2000). At smaller $\lambda$ and $Re_\infty$ the deviations are $< 10\%$, in some cases much smaller. The larger deviations (~21%, 62%) in the two smallest $\lambda$ entries in Table 11 arise from anomalous values of $f^V$ (sec. 4.4.8). In tubes with $D = 15$ the relative deviations are: ~7-16% and only for the two largest $\lambda$ are ~50% and 180% respectively ($Re_\infty$ = 3910 and 7105 respectively); in $D = 30 \, mm$, the relative deviations are: 17-34% for the first four entries and ~80% for the largest $\lambda$ ($Re_\infty$ = 7300). The values of $\lambda$ for which large $Re_\infty$ and large $\Delta f^\mu / f^\mu$ and $\Delta \mu_\infty / \mu_\infty$ are observed (Table 11) are considerably lower ($\lambda$~0.15) than the value range of error-prone $\lambda$ ($> 0.4$) in the smaller $Re_\infty$ range ($<50$). $\Delta f^\mu$ for large $Re_\infty$ is smaller than that for smaller $Re_\infty$, but so is $f^\mu$; the ratio in some cases is larger for large $Re_\infty$.

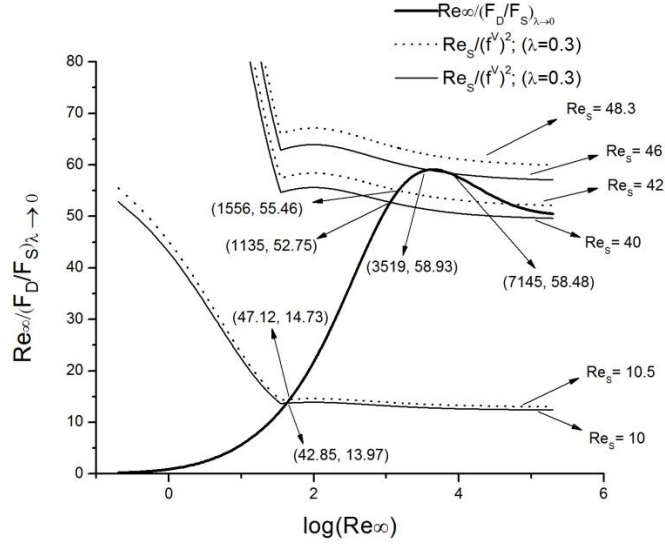

**Figure 14: Sensitivity of $Re_\infty$ to $\Delta V_t$.**

Clearly the problem is reduced if $Re_\infty$ can be lowered at a given $\lambda$. It is possible to have a handle on $Re_\infty$ by lowering the ball material density. In Fig. 5b plots of $Re_\infty$ vs. $d$ for several values of $\Delta\rho$ (within a range of easily available ball materials) show that a suitable choice of $d$ and $\rho_b$ can reduce $Re_\infty$ by a factor of 3 or 4. Polystyrene balls which are less dense than the ball materials we consider can reduce the problem further, but they may pose problems with some denser liquids namely Glycerol. Since liquid densities vary over a narrow range, our conclusions based on specific liquid densities will generally hold. Design of problem specific $\rho_b$ can reduce $Re_\infty$, bring it within the creepy flow regime and can simplify analysis.

**Table 13: Propogation of error in $V_t$ to that in $\mu_\infty$ for various $\alpha$ values.**

| $Re_\infty$ | $\alpha$ | $\alpha^*$ | $\dfrac{\Delta(F_D/F_S)_{\lambda\to 0}}{(F_D/F_S)_{\lambda\to 0}}$ $\dfrac{\Delta V_t}{V_t}=0.5\%$ | $\dfrac{\Delta\mu_\infty}{\mu_\infty}$ | $\alpha^*$ | $\dfrac{\Delta(F_D/F_S)_{\lambda\to 0}}{(F_D/F_S)_{\lambda\to 0}}$ $\dfrac{\Delta V_t}{V_t}=5\%$ | $\dfrac{\Delta\mu_\infty}{\mu_\infty}$ |
|---|---|---|---|---|---|---|---|
| 25 | 10.10 | 10.20 | -0.67% | 0.17% | 11.11 | -6.69% | 1.59% |
| 100 | 21.77 | 21.99 | -0.98% | 0.48% | 23.95 | -10.15% | 4.67% |
| 500 | 42.90 | 43.33 | -2.18% | 1.64% | 47.19 | -25.94% | 16.62% |
| 1000 | 51.46 | 51.98 | -4.02% | 3.38% | 56.61 | -67.89% | 37.46% |
| 2500 | 58.09 | 58.67 | -20.18% | 16.38% | 63.89 | | |
| 10000 | 57.59 | 58.16 | 19.28% | -24.50% | 63.35 | | |

$$\alpha^* = \alpha\left(1 + 2\frac{\Delta V_t}{V_t}\right)$$

## Conclusion

It is possible to extend the use of falling ball viscometry with excellent accuracy well beyond the 'creepy flow' regime with use of the method described in this paper.

## Acknowledgement

The work is based on the Master's theses of A. V. Singh and L. Sharma and subsequent work done by them as Research Fellows at IIT Kanpur. They thank IIT Kanpur for the financial assistance. The authors also thank Debiprasad Palit, Asoke Kumar Mallik, Vishal Saxena, Aloke Trivedi, S. Vishnu, Mitesh, Naresh, Saurabh Paul, Subhrajyoti Adhya, V. Chandrasekhar, S. G. Dhande, R. N. Mukherjee, P.K. Bhardwaj, , Salman Khan, R. S. Rajput, P. H. Tewari, Piyush Sahay, Sanghamitra Deb, Arunava Ghosh and. Anil Kumar Verma